\documentclass[iop]{emulateapj}

\usepackage{braket}

\shorttitle{}
\shortauthors{Hasegawa et al}

\begin{document}

\title{Magnetically Induced Disk Winds and Transport in the HL Tau Disk}

\author{Yasuhiro Hasegawa\altaffilmark{1}, Satoshi Okuzumi\altaffilmark{2}, Mario Flock\altaffilmark{1}, and Neal J. Turner\altaffilmark{1}}
\affil{$^1$Jet Propulsion Laboratory, California Institute of Technology, Pasadena, CA 91109, USA}
\affil{$^2$Department of Earth and Planetary Sciences, Tokyo Institute of Technology, Meguro-ku, Tokyo 152-8551, Japan}
\email{yasuhiro@caltech.edu}

\begin{abstract}
The mechanism of angular momentum transport in protoplanetary disks is fundamental to understand the distributions of gas and dust in the disks.
The unprecedented, high spatial resolution ALMA observations taken toward HL Tau and subsequent radiative transfer modeling reveal that
a high degree of dust settling is currently achieved at the outer part of the HL Tau disk.
Previous observations however suggest a high disk accretion rate onto the central star.
This configuration is not necessarily intuitive in the framework of the conventional viscous disk model, 
since efficient accretion generally requires a high level of turbulence, which can suppress dust settling considerably.
We develop a simplified, semi-analytical disk model to examine under what condition these two properties can be realized in a single model.
Recent, non-ideal MHD simulations are utilized 
to realistically model the angular momentum transport both radially via MHD turbulence and vertically via magnetically induced disk winds.
We find that the HL Tau disk configuration can be reproduced well when disk winds are properly taken into account.
While the resulting disk properties are likely consistent with other observational results, 
such an ideal situation can be established only if the plasma $\beta$ at the disk midplane is $\beta_0 \simeq 2 \times 10^4$
under the assumption of steady accretion.
Equivalently, the vertical magnetic flux at 100 au is about 0.2 mG.
More detailed modeling is needed 
to fully identify the origin of the disk accretion and quantitatively examine plausible mechanisms behind the observed gap structures in the HL Tau disk.
\end{abstract}


\section{Introduction} \label{sec:intro}

Protoplanetary disks are the birthplace of planetary systems including the solar system.
The spatial distributions of both gas and dust in the disks provide the initial conditions for planet formation. 
The time evolution of the disks dictates when and where planet formation begins, 
and eventually determines the final architecture of planetary systems both in mass and in orbital distance \citep[e.g.,][]{il04i,mab09,bia14,h16,mvm16}.
Such distributions of planets can be confronted with a wealth of observed exoplanetary populations \citep[e.g.,][]{us07,mml11,bkb11,wf15,tjs16}.
 
It is widely perceived that magnetic fields threading disks can regulate the evolution \citep[e.g.,][]{a11}.
While the ultimate mechanism of the observed high disk accretion rate is still inconclusive \citep[e.g.,][]{hhc16},
magnetorotational instability (MRI) and the resulting MHD turbulence can act as a promising candidate to account for the observations \citep[e.g.,][]{bh98}.
This becomes possible because MRI can efficiently transport the disk angular momentum radially 
by coupling magnetic fields with differentially rotating ionized gas in disks.
One of the supportive evidence for the presence of magnetic fields in disks may be magnetized chondrules found in chondrites \citep{fwl14}.
The results of the lab experiments imply that magnetic fields should have played an important role 
in both the distribution and the growth of planet-forming materials in the solar nebula \citep[e.g.,][]{ssl96,dc00,c07b,htm16,wwb17}.

The MRI and the resulting MHD turbulence do not necessarily fully operate in protoplanetary disks, 
since the disks are generally dense and cold in most regions \citep[e.g.,][]{nfg14}.
Accordingly, non-ideal MHD effects become important to realistically model a coupling between charged materials and magnetic fields in disks 
and to reliably compute the disk angular momentum transport.
Ohmic resistivity has been investigated extensively over the past two decades.
It is known that inclusion of Ohmic resistivity leads to the realization of layered accretion \citep[e.g.,][]{g96,alp01,fs03,t08,kl10,ht15}.
In the picture, the disk angular momentum is transported radially via MHD turbulence; a higher disk accretion rate is achieved in the MRI-active surface layers
while in the disk midplane, MRI is quenched and MRI-dead zones are present,
which ends up with a lower disk accretion rate there.
Compared with Ohmic resistivity, other non-ideal MHD terms (ambipolar diffusion and the Hall term) have been less explored \citep[e.g.,][]{w99,ss02}.
The situation has been rapidly changing immediately after the recognition that 
ambipolar diffusion can suppress MRI in the disk surfaces significantly \citep[e.g.,][]{bs11}.
For this case, the observed high disk accretion rate can be reproduced only when disks are threaded by strong magnetic fields \citep[e.g.,][]{sba13}.
This condition is needed 
because then magnetically induced disk winds can be launched from the disk surface and efficiently transport the disk angular momentum vertically \citep[e.g.,][]{si09,bs13,sba13}.
Inclusion of different non-ideal MHD terms thus draws a very different picture of how protoplanetary disks evolve with time \citep[e.g.,][]{b16,som16}:
depending on the strength of the vertical magnetic fields,
the disk angular momentum can be possibly transported radially via the MRI and the resulting MHD turbulence 
and/or vertically via magnetically induced disk winds.

Observations of protoplanetary disks serve as an important complementary approach 
to understand disk evolution and planet formation there \citep[e.g.,][]{wc11}.
The advent of ALMA with significantly upgraded sensitivity and angular resolution has indeed revolutionized our view of planet formation \citep[e.g.,][]{fvc15}.
The best example is the observation taken toward HL Tau as a part of the ALMA long-baseline science verification campaign \citep{bph15}.
The astonishing observations reveal that nearly concentric, multiple gaps are present in the dust continuum emission in the disk,
which may be a potential signature of planet formation in a young stellar object (YSO).
It is remarkable that similar gap structures in disks have recently been reported for a number of other targets 
\citep[e.g.,][]{ntk16,awz16,tnm16,igt16,fch17}.
Soon after the public release of the HL Tau image,
a number of mechanisms have been proposed to explain the origin of gaps for the HL Tau disk.
These include disk-planet interaction \citep[e.g.,][]{dpl15,dzw15,kmt15,ahh16,jli16}, the effect of ice lines \citep{zbb15}, sintering effects for growing dust particles  \citep{oms16},
and secular gravitational instabilities \citep{ti16}.

While all the studies can (partially) reproduce the observed gap structure for the HL Tau disk,
the predominant mechanism of such local structures is not fully identified yet.
This occurs probably because different mechanisms are examined in different disk models,
and hence a direct comparison among them may not be trivial. 
Furthermore, it is evident based on the Br$\gamma$ observations that 
the HL Tau disk has a high accretion rate \citep[$\simeq 10^{-7}$ M$_{\odot}$ yr$^{-1}$,][]{bbm10}.
This is intuitively expected since HL Tau is a relatively young system ($\la 1$ Myr).
Most importantly, radiative transfer modeling of the ALMA data indicates millimeter-sized grains have settled greatly at $\sim 100$ au from the star, 
reducing their scale height to just $\sim 1$ au \citep{pdm16}.
Turbulence at this location must then be weak.  
Further constraints on the turbulent strength come from the widths of the millimeter-wave emission lines of several gas molecules
\citep{hwa11,gdw12,fhr15,fhr17}, 
though concerns have been raised about the measurements' sensitivity to uncertainties in the gas temperature structure and flux calibration \citep{tgs16}.
Weak turbulence may be incompatible with the conventional viscous disk model, 
because rapid accretion requires strong turbulence, which counteracts the grains' settling.  
Thus, it is worth examining the HL Tau disk's global configuration before addressing the origin of the observed gaps.

In this paper, we build such a global view, especially focusing on the disk accretion rate and vertical mixing of dust particles by disk turbulence.
We develop a semi-analytical disk model, making use of recent non-ideal MHD simulations.
The key ingredient in our model is to properly include the effects of both MHD turbulence and disk winds simultaneously.
We will show below that under the framework of magnetically driven disk accretion,
the two observational features (a high disk accretion rate and efficient dust settling) can be reproduced in single disk models 
when the plasma $\beta$ at the disk midplane is $\beta_0 \simeq 2 \times 10^4$
under the steady accretion assumption.

The plan of this paper is as follows.
In Section \ref{sec:mod}, we develop a simplified, but physically motivated disk model, using the results of recent non-ideal MHD simulations.
In Section \ref{sec:res}, we present our results and examine how the observed properties of the HL Tau disk can be reproduced in our model.
We also perform a parameter study and consider applications of our results.
These include the global configuration of magnetic fields and the gas-to-dust ratio.
In Section \ref{sec:disc}, we take into account other observational features of the HL Tau disk.
We also discuss limitations of our models and other potential mechanisms of disk accretion for the HL Tau system.
Our conclusions are provided in Section \ref{sec:conc}.

\section{Disk model} \label{sec:mod}

\subsection{Magnetically driven disk accretion}

When magnetic fields play an important role in transporting angular momentum in disks,
the disk accretion rate ($\dot{M}$) can be written as \citep[e.g.,][]{fll13,som16}
\begin{eqnarray}
\dot{M} & = & \frac{4\pi}{r\Omega}
\label{eq:mdot_b}  
                       \left\{   \frac{\partial }{\partial r}
                       \left[   r^2 \int_{-H_{\rm w}}^{H_{\rm w}}  dz  \left( \overline{ \rho v_r \delta v_\phi } 
                                                                                               - \frac{ \overline{ B_r B_\phi } }{ 4\pi }    \right) 
                        \right] \right. \nonumber \\
             &       & \left. +  r^2  \left( \overline{ \rho v_z \delta v_\phi } 
                                                     - \frac{ \overline{ B_z B_\phi } }{4\pi}   \right)_{z=-H_{\rm w}}^{z=H_{\rm w}}
                       \right\},
\end{eqnarray}
where the cylindrical coordinate system is adopted.
The conventional notation is used here; $r$ is the disk radius from the central star, $\Omega = \sqrt{GM_*/r^3}$ is the orbital frequency,
$M_*$ is the mass of the central star, $\rho$ is the gas volume density, 
and $v_i$ and $B_i$ are the $i-$th component of the gas velocity and of magnetic fields, respectively.
Also, $\delta v_{\phi}$ is computed as $v_{\phi}- r \Omega$, 
and the time-averaged value is utilized for some of physical quantities with the notation of the overbar.
In addition, it is assumed in the equation that 
disk surfaces are located at $z = \pm H_{\rm w}$ from the midplane, and disk winds will be launched from these surfaces.
As discussed below (see Section \ref{sec:mod_fit}), $H_{\rm w} \simeq 2H_{\rm g}$, where $H_{\rm g}=c_{\rm s} / \Omega$ is the pressure scale height and $c_{\rm s}$ is the sound speed.
This equation is obtained from a vertical integration of the azimuthal component of the MHD momentum equation over $|z| < H_{\rm w}$, 
and is applicable to both the cases of non-steady and steady disk accretion.

We introduce the following notation for the normalized accretion stresses,
which are written as
\begin{equation}
W_{r\phi} \equiv \frac{ \int_{-H_{\rm w}}^{H_{\rm w}} \left(\overline{\rho v_r\delta v_\phi} - \overline{B_rB_\phi}/4\pi \right) dz}{\Sigma_{\rm g} c_{\rm s}^2},
\end{equation}
and
\begin{equation}
W_{z\phi} \equiv \frac{ \left(\overline{\rho v_z\delta v_\phi} - \overline{B_zB_\phi}/4\pi \right)_{z=-H_{\rm w}}^{z=H_{\rm w}}}{2\rho_{\rm mid}c_{\rm s}^2},
\end{equation}
where $\Sigma_{\rm g}$ is the gas surface density, and $\rho_{\rm mid} = \Sigma_{\rm g} \Omega/(\sqrt{2\pi} c_{\rm s})$ is the gas volume density at the disk midplane.
The vertically isothermal assumption is adopted for $c_{\rm s}$.
We here follow the notation used in \citet{sba13}; our $W_{r\phi}$ and $W_{z\phi}$ correspond to their $\alpha_{\rm bw}$ and $|W_{z\phi}|_{\rm bw}$, respectively.
Note that the normalization constants are different for the above two equations.
This is because $W_{r\phi}$ is the accretion stress integrated for the vertical direction,
while the wind stress ($W_{z\phi}$) can be regarded as the excess of the angular momentum flux coming into/out of the disk surfaces.
This difference is properly taken into account when computing the disk accretion rate (see a factor of $2r/(\sqrt{\pi}H_{\rm g})$ in equation (\ref{eq:mdot_b_s})).
Then, equation (\ref{eq:mdot_b}) can be re-written as
\begin{equation}
\label{eq:mdot_b2}
\dot{M} = \frac{4\pi}{r\Omega}\frac{\partial}{\partial r}\left(r^2 c_{\rm s}^2 \Sigma_{\rm g} W_{r\phi} \right)
+ 4\sqrt{2\pi}r \Sigma_{\rm g} c_{\rm s} W_{z\phi}.
\end{equation}

Thus, magnetic fields can regulate disk evolution by transporting the disk angular momentum 
both radially via MHD turbulence ($W_{r\phi}$) and vertically via magnetically induced disk winds ($W_{z\phi}$).

\subsection{Conventional viscous disk models}

The classical 1D viscous disk model is still widely adopted in the literature to model protoplanetary disks.
In this picture, it is assumed that  the radial angular momentum transport is determined by a kinematic viscosity ($\nu$) without specifying its origin,
and the disk accretion rate can be written as \citep[e.g.,][]{p81}
\begin{equation}
 \label{eq:mdot_vis}
 \dot{M} = \frac{6\pi }{\Omega r}\frac{\partial}{\partial r}\left(r^2\Omega\nu \Sigma_{\rm g} \right).
\end{equation}

We can obtain a relation between the accretion stress ($W_{r\phi}$) and viscosity ($\nu$) 
by comparing equation (\ref{eq:mdot_vis}) with the first term in the right-hand side of equation (\ref{eq:mdot_b2}):
\begin{equation}
W_{r\phi} = \frac{3}{2} \frac{ \nu \Omega }{c_{\rm s}^2}.
\end{equation}
This follows that the $\alpha$ viscous parameter ($\alpha_{\rm SS}$) can be expressed as a function of $W_{r\phi}$ \citep{ss73},
which is given as
\begin{equation}
\label{eq:alpha_ss}
\alpha_{\rm SS} = \frac{2}{3} W_{r\phi},
\end{equation} 
where $\alpha_{\rm SS} \equiv \nu \Omega / c_{\rm s}^2$.
Note that the factor of 2/3 is often neglected in the literature.

Since equation (\ref{eq:mdot_vis}) only considers the radial angular momentum transport, 
it would work well for modeling the accretion rate and the resulting disk structure
when turbulence plays a dominant role in disk evolution.
In other words, disk winds should not transport the angular momentum efficiently 
either due to weak vertical magnetic flux or due to an unfavorable geometrical configuration of the magnetic field.

\subsection{Steady state solutions}

We explore steady disk accretion ($\dot{M} = {\rm const.}$) using equation (\ref{eq:mdot_b2}).
While a more general solution can be obtained by integrating equation (\ref{eq:mdot_b2}) along the radial direction,
we here derive an approximate but analytic solution by examining two limiting cases of disk accretion. 

First, we consider the limit where disk accretion is dominated by the internal disk stress ($W_{r\phi}$).
In this limit, the steady state condition ($\dot{M} = {\rm const.}$) can be realized
when the product $c_{\rm s}^2 \Sigma_{\rm g} W_{r\phi}$ is proportional to $\Omega (\propto r^{-3/2})$.
Then, $\partial (r^2 c_{\rm s}^2 \Sigma_{\rm g} W_{r\phi}) / \partial r  = (1/2)r c_{\rm s}^2 \Sigma_{\rm g} W_{r\phi}$.
Accordingly, we find that the disk accretion rate can be written as 
\begin{equation}
\dot{M} \approx \frac{2\pi \Sigma_{\rm g}  c_{\rm s}^2 W_{r\phi}}{\Omega}.
\end{equation}
The other limit is that disk accretion is regulated predominantly by the wind stress, that is, $\dot{M} \approx 4\sqrt{2\pi}r \Sigma_{\rm g} c_{\rm s} W_{z\phi}$.

Second, we combine the solutions of the above two limits to consider a more general situation.
For this case, the steady disk accretion rate can be approximately given as 
\begin{eqnarray}
\label{eq:mdot_b_s}
\dot{M} &\approx& \frac{2\pi \Sigma_{\rm g} c_{\rm s}^2 W_{r\phi}}{\Omega} + 4\sqrt{2\pi}r \Sigma_{\rm g} c_{\rm s} W_{z\phi}  \nonumber \\
            & =           & \frac{2\pi \Sigma_{\rm g} c_{\rm s}^2}{\Omega} \left( W_{r\phi} +  \frac{ 2r }{ \sqrt{\pi}H_{\rm g} } W_{z\phi} \right).
\end{eqnarray}

Once $\dot{M}$ is obtained, 
we can also compute both the surface density profile ($\Sigma_{\rm g}$) and the effective $\alpha-$parameter ($\alpha_{\rm SS, eff} $) 
for disks with $\dot{M} = {\rm const.}$;
\begin{equation}
\Sigma_{\rm g} = \frac{\dot{M}\Omega}{2\pi c_{\rm s}^2}\left( W_{r\phi} +  \frac{ 2r }{ \sqrt{\pi}H_{\rm g} } W_{z\phi} \right)^{-1},
\end{equation}
and 
\begin{equation}
\label{eq:alpha_ss_eff}
\alpha_{\rm SS, eff} = \frac{2}{3} \left( W_{r\phi} + \frac{ 2r }{ \sqrt{\pi}H_{\rm g} } W_{z\phi} \right).
\end{equation}
Note that $\alpha_{\rm SS, eff}$ is derived from the comparison between equation (\ref{eq:mdot_b_s}) and the disk accretion rate in the 1D viscous model,
which can be given as (see equation (\ref{eq:mdot_vis}))
\begin{equation}
\label{eq:mdot_vis_s}
\dot{M} = 3\pi \alpha_{\rm SS} c_{\rm s}^2 \Sigma_{\rm g} /\Omega.
\end{equation}

Thus, disk winds provide an additional contribution to $\dot{M}$, $\Sigma_{\rm g}$, and the $\alpha-$parameter. 
Furthermore, since the contribution is scaled as $r / H_{\rm g}$,
they can play a dominant role in the evolution when disks are threaded by magnetic fields 
that have the corresponding plasma $\beta_0$ of $10^{3}$ or lower at the disk midplane (see Figure \ref{fig1}).

\subsection{Dust settling}

Dust settling is one of the key processes to infer the level of disk turbulence in protoplanetary disks \citep[e.g.,][]{md12}.
This is because the degree of dust settling is regulated both by the dust particle size and by disk turbulence \citep[e.g.,][]{dms95,yl07}.
Here we briefly summarize the formulation of dust settling that is used in this paper.

The vertical diffusion coefficient for small dust particles can be defined as \citep[e.g.,][]{zsb15}
\begin{equation}
D_z = \frac{1}{2}\frac{d\langle z^2\rangle}{dt},
\end{equation}
where the brackets denote the ensemble average taken over the trajectories of a large number of particles.
Then, the normalized diffusion coefficient can be written as 
\begin{equation}
\label{eq:alpha_D}
\alpha_{D} \equiv \frac{D_z}{c_{\rm s}H_{\rm g}}.
\end{equation}
In general, dust particles with the radius of $a$ ($< 1-10$ mm) in protoplanetary disks can reside in the so-called Epstein regime
where $a$ is smaller than the mean free path of gas particles.
For this case, the Stokes number that is the dimensionless stopping time 
can be given as ${\rm St=(\pi \rho_s a)/(2\Sigma_{\rm g})}$ for such particles \citep[e.g.,][]{bke12}.
Finally, the scale height ($H_{\rm d}$) of the dust particles can be written as \citep[e.g.,][]{yl07}
\begin{equation}
 \label{eq:H_d}
 H_{\rm d} = \left(1+ \frac{{\rm St}}{\alpha_{D}}\right)^{-1/2} H_{\rm g}.
\end{equation}
This is the outcome that dust settling equilibrates with the vertical stirring due to disk turbulence.
Note that $\alpha_D$ needs not to be the same as $\alpha_{\rm SS}$ and $\alpha_{\rm SS,eff}$.

\subsection{Fitting formulae for $W_{r\phi}$, $W_{z,\phi}$ and $\alpha_D$} \label{sec:mod_fit}

\begin{figure*}
\begin{minipage}{17cm}
%\begin{figure}%[!ht]
\begin{center}
\includegraphics[width=8cm]{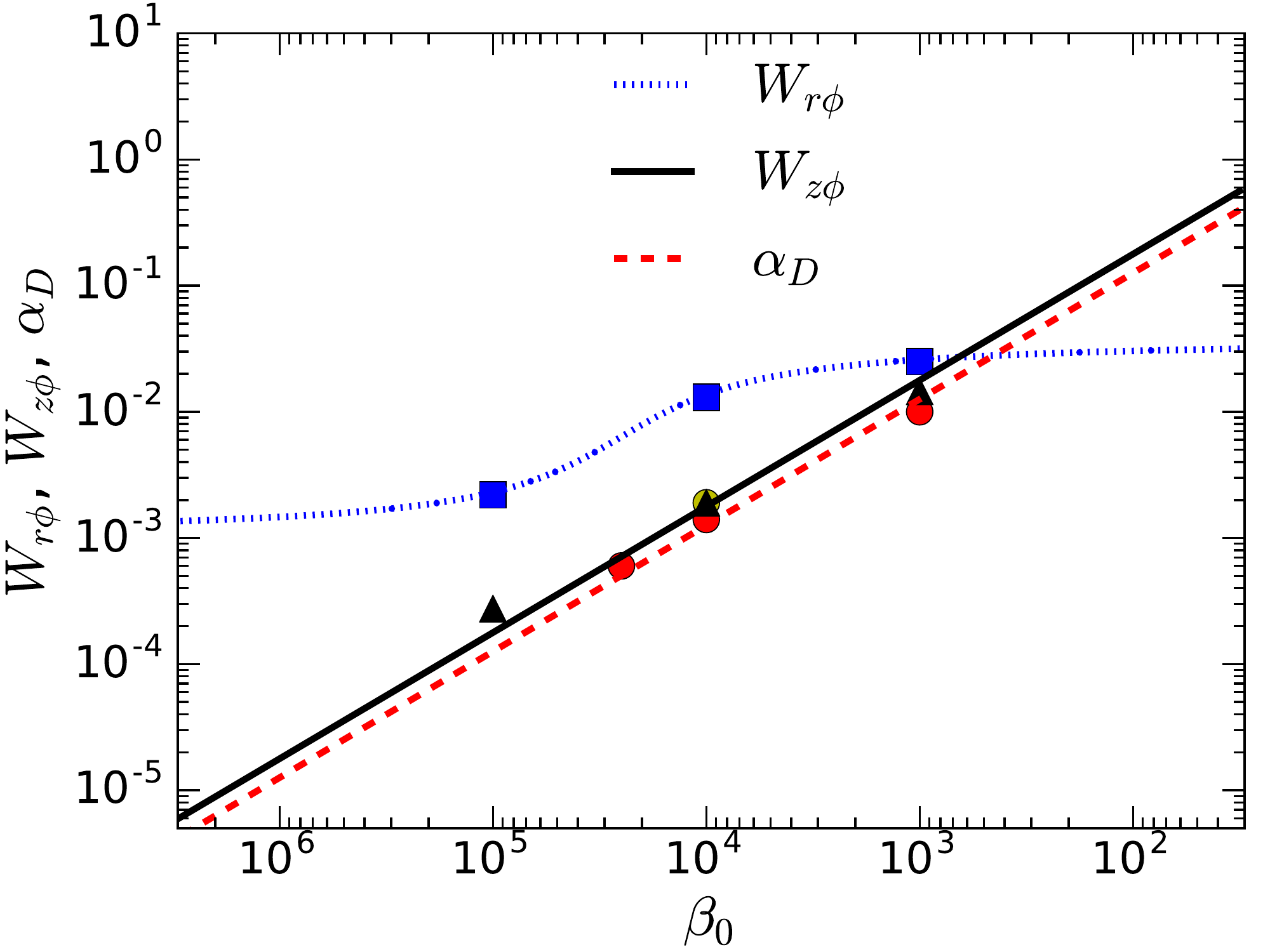}
\includegraphics[width=8cm]{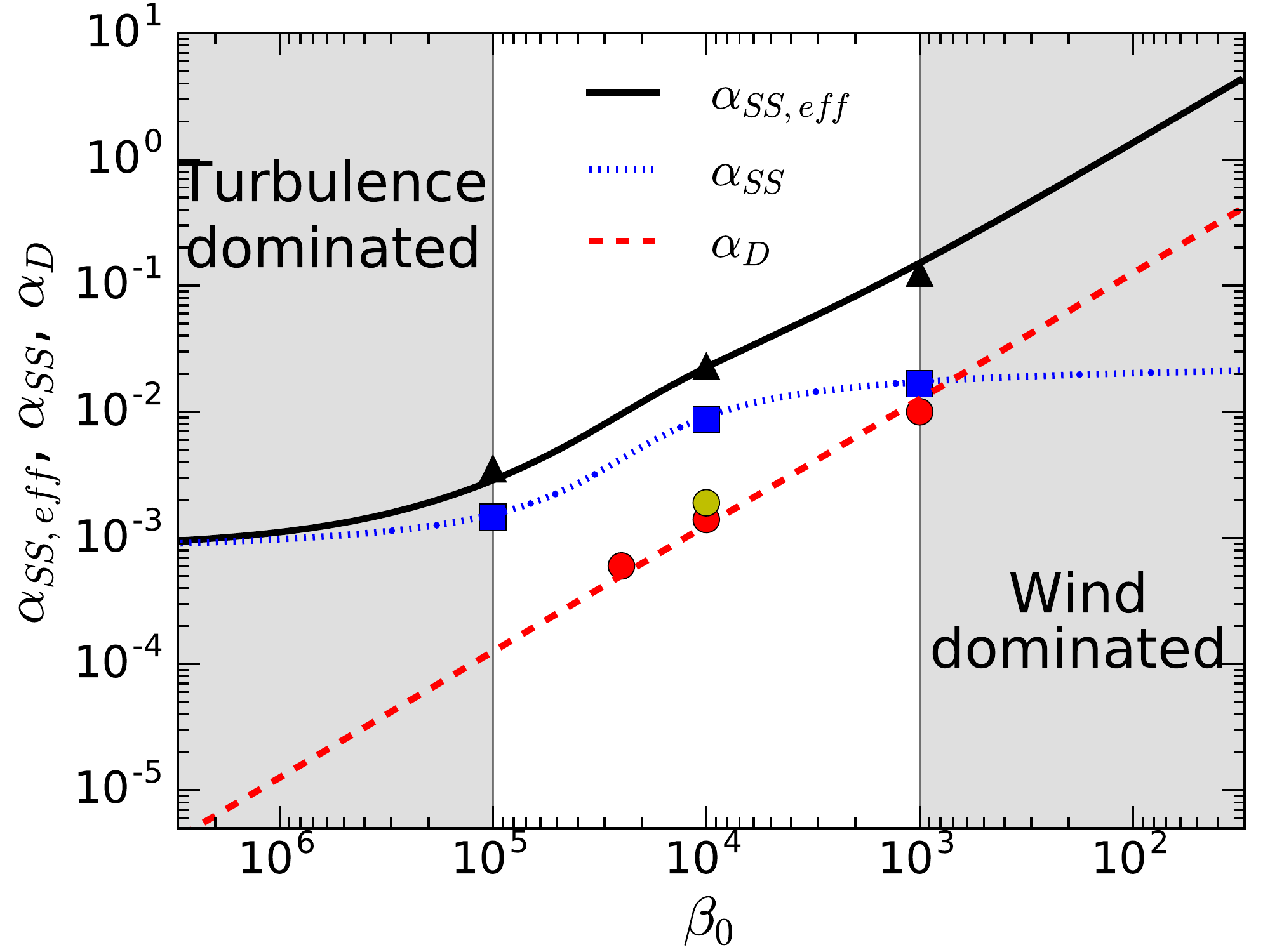}
\caption{Dimensionless stresses as a function of $\beta_0$. 
On the left, the fitting formulae for $W_{r\phi}$, $W_{z\phi}$, and $\alpha_D$ are present with the results of numerical simulations.
The blue squares and the black triangles are adopted from the results of \citet{sba13} (see their models AD30AU1e3 - AD30AU1e5)
while the red circles and the yellow one are from \citet{zsb15} (see their models AD1R64 - AD3R64 and AD2SR32).
Our fitting formulae for these stresses work well for given values of $\beta_0$.
On the right, the resulting $\alpha_{\rm SS}$ and $\alpha_{\rm SS,eff}$ are shown (see equations (\ref{eq:alpha_ss}) and (\ref{eq:alpha_ss_eff})).
Here, we set that $r/H_{\rm g}=10$ as an example.
When $\beta_0$ is large, MHD turbulence is important for $\alpha_{\rm SS, eff}$ (see the turbulence dominated regime).
As $\beta_0$ decreases, that is, the initial vertical magnetic field becomes stronger,
magnetically induced disk winds become more dominant (see the wind dominated regime).}
\label{fig1}
\end{center}
%\end{figure}
\end{minipage}
\end{figure*}

Based on the above formalism, 3 quantities ($W_{r\phi}$, $W_{z,\phi}$ and $\alpha_D$) are needed 
to uniquely determine the spatial distributions of both gas and dust for a given disk temperature profile (equations (\ref{eq:mdot_b_s}) and (\ref{eq:H_d})).

Here, we make use of recent non-ideal MHD simulations to model these quantities as a function of the plasma $\beta$,
where 
\begin{equation}
\label{eq:beta}
\beta = \frac{8 \pi \rho_{\rm mid} c_{\rm s}^2}{ B_z^2}.
\end{equation}
More specifically, the simulation results obtained by \citet{sba13} and \citet{zsb15} are utilized,
which are both appropriate for the outer part of protoplanetary disks.
\citet{sba13} perform vertically stratified, local shearing box, 3D MHD simulations with ambipolar diffusion.
In these simulations, the values of $W_{r\phi}$ and $W_{z\phi}$ are measured at the disk height of the wind base ($z = \pm z_{\rm bw}$) 
for three different values of the initial $\beta$ ($\beta_0 = 10^3$, $10^4$, and $10^5$, see Table 1 of \citet{sba13}).
They find that $z_{\rm bw} \simeq 2 H_{\rm g}$ for a wide range of $\beta_0$. 
In this paper, it is assumed that $H_{\rm w} \equiv z_{\rm bw}$ (see equation (\ref{eq:mdot_b})).
For the actual fitting, we follow the approach taken by \citet{asm13},
where simple fitting formulae are derived to reproduce the simulation results;
\begin{equation}
 \log W_{r\phi} = -2.2 + 0.5 \tan^{-1}\left(\frac{4.4-\log \beta_0}{0.5}\right),
\end{equation}
and
\begin{equation}
\label{eq:W_zphi}
 \log W_{z\phi} = 1.25 - \log \beta_0.
\end{equation}
While the formula for $W_{z\phi}$ is identical to that used in \citet{asm13},
our formula for $W_{r\phi}$ is slightly different from theirs 
because we here consider the internal disk stress averaged over $|z| \leq z_{\rm bw}$, rather than  $|z| \leq 4H_{\rm g}$.

For $\alpha_D$, the results of \citet{zsb15} are adopted,
wherein both unstratified and stratified 3D local shearing box, non-ideal MHD simulations with Lagrangian particles are performed.
In these simulations, ambipolar diffusion is taken into account, 
and the values of $\alpha_D$ are measured over $|z| \leq H_{\rm g}$ for different values of the initial $\beta$.
Based on the results of their unstratified simulations (see their Table 3),
we find that $\alpha_{D}$ is roughly proportional to $1/\beta_0$.
Mathematically, the resultant fitting formula can be expressed as     
\begin{equation}
\log \alpha_{D} = 1.1-\log \beta_0.
\end{equation}

Figure \ref{fig1} shows the fitting profiles for $W_{r\phi}$, $W_{z\phi}$, and $\alpha_D$ and 
the resulting $\alpha_{\rm SS,eff}$ and $\alpha_{\rm SS}$ on the left and the right panels, respectively.
As an example, we adopt that $r/H_{\rm g}=10$ when computing $\alpha_{\rm SS, eff}$ in this figure (see equation (\ref{eq:alpha_ss_eff})).
In addition, the numerical results utilized for fitting are also plotted;
the blue squares and the black triangles are adopted from \citet{sba13}.
The red circles and the yellow circle are from the results of unstratified and stratified simulations done by \citet{zsb15},
respectively.
Figure \ref{fig1} shows that our formulae can fit the results of the numerical simulations very well (see the left panel).
While it is out of the scope of this paper, it is interesting that $\alpha_D$ tends to follow $W_{z\phi}$ more than $W_{r\phi}$.
Some discussion about the ratio between $W_{r\phi}$ and $\alpha_D$, which is also known as the schmidt number, can be found in \citet{zsb15}.
Also, our fitting suggests (the right panel) that 
MHD turbulence is significant for the accretion stress ($\alpha_{\rm SS, eff}$) when $\beta_0$ is large (see the turbulent dominated regime).
As $\beta_0$ decreases, disk winds play a more crucial role in disk evolution (see the wind dominated regime).

Note that care is needed in using these fitting formulae for the regimes of $\beta_0 > 10^5$ and $\beta_0 < 10^3$.
For the case of $\beta_0 > 10^5$, our $W_{r\phi}$ approaches the constant value of $\sim 10^{-3}$, 
and both of $W_{z\phi}$ and $\alpha_D$ decrease linearly with increasing $\beta_0$.
The behavior of $W_{r\phi}$ is supported by non-ideal MHD simulations from \citet{sbs13} which show that
the value of $W_{r\phi}$ is on the order of $10^{-3}$ for the case of no vertical magnetic fields (equivalently $\beta_0 = \infty$).
For $W_{z\phi}$ and $\alpha_D$, there is currently no simulation to examine these behaviors.
In the regime of $\beta_0 < 10^3$, our formulae suggest that $W_{r\phi}$ saturates to $\simeq 10^{-2}$
while both $W_{z\phi}$ and $\alpha_D$ keep increasing.
As with the above case, the results of non-ideal MHD simulation are not available in the literature,
so that the behavior of these three quantities cannot be examined.
(Even ideal MHD simulations are limited for the case of $\beta_0 < 10^3$ \citep[e.g.,][]{shg96,bs13a}.)
Likely $W_{r\phi}$ drops to zero as $\beta_0$ decreases,
since strong vertical magnetic fields can fully quench MRI and hence suppress MHD turbulence \citep[e.g.,][]{ksw10}.

Thus, both the accretion stress ($\alpha_{\rm SS, eff}$) and the dust diffusion coefficient ($\alpha_{D}$) are evaluated simultaneously 
for a given value of $\beta_0$, yielding reasonable values for $10^5 > \beta_0 > 10^3$.
This makes it possible to examine in our framework how the global structure of disks is determined as a function of $\beta_0$ 
and how different the vertical distribution of dust particles is for different values of $\beta_0$.

\subsection{Model parameters for the disk around HL Tau} \label{sec:model_kwon}

We finally summarize the key properties of the disk around HL Tau.
These properties are derived either only from observations or 
from comparisons between observations and radiative transfer modeling.

We assume that $M_* = 1 M_{\odot}$ and $\dot{M} =  1 \times 10^{-7}$ $M_{\odot}$ yr$^{-1}$ for the HL Tau system.
These two values are chosen because they are in the range suggested by observations;
$0.5 \la M_{*}/M_{\odot} \la 1.3$ \citep{klm11,bph15}\footnote{Recently, \citet{pdm16} have found that 
the HCO$^+$ and CO line emission from the HL Tau disk are consistent with Keplerian motion around a star with $1.7 M_{\odot}$, 
although HL Tau is classified as a K-type star.} 
and $0.9 \times 10^{-7} \la \dot{M}_{\rm g} /(M_{\odot} {\rm yr}^{-1}) \la 2 \times 10^{-6}$ \citep{hom93,bbm10}.
For the dust surface density distribution ($\Sigma_{\rm d}$) and the disk temperature ($T_{\rm mid}$), we follow \citet{klm11}.
In their model, similarity solutions are adopted for $\Sigma_{\rm d}$, and a simple power-law profile is used for $T_{\rm mid}$,
which are given as
\begin{eqnarray}
\label{eq:sigma_d}
\Sigma_{\rm d} & = & \rho_{d0} \sqrt{2 \pi} H_{\rm g} \left( \frac{r}{r_c} \right)^{-p} \exp \left[ - \left( \frac{r}{r_c} \right)^{7/2-p-q/2} \right], \\
T_{\rm mid} & = & T_0 (r/r_0)^{-q},
\end{eqnarray}
where $r_0=10$ au, $q=0.43$, and $T_0=62$ K.
They find that the observed surface brightness of the disk around HL Tau can be reproduced very well 
when $\rho_{d0} \simeq 2.2 \times 10^{-15}$ g cm$^{-3}$, $r_c = 78.9$ au, and $p \simeq 1$,
which are confirmed by their latest work \citep{klm15}.
Note that the above model parameters are obtained from a comparison with the modest resolution observations done by CARMA \citep{klm11,klm15}.
The recent studies suggest that except for gap regions,
Kwon's disk model works well both for ALMA observations with very high spatial resolutions \citep{ahh16,pdm16}
and for JVLA observations with comparable resolutions \citep{chc16}.

Another key result is that the midplane layer rich in millimeter-sized dust particles is surprisingly thin 
with the scale height of $H_{\rm d} \approx 1$ au at $r \approx 100$ au. 
This follows from the fact that the gaps between the bright rings are clearly visible along the projected minor axis, despite the system's inclination to our line of sight.
The resulting scale height estimate comes from detailed radiative transfer calculations \citep{pdm16}.

In the following, we use the above parameters to compute the disk structure around HL Tau.

\section{Results \& Applications} \label{sec:res}

We present the results that are derived from magnetically-driven disk accretion model.
We also conduct a parameter study to examine how our results are maintained for a wide range of model parameters.
Furthermore, we discuss the applications of our results such as the global structure of magnetic fields and the gas-to-dust ratio in the outer part of the HL Tau disk.

\subsection{Results from disk wind models}

We first present the gas and dust structures coming from the magnetically-driven disk accretion model.  
Substituting the prescriptions for the turbulent and wind stresses in equation (\ref{eq:mdot_b_s}) yields the surface density profile ($\Sigma_{\rm g}$) 
for a given value of the net vertical flux's midplane plasma beta ($\beta_0$).  
In turn, substituting the prescription for the turbulent diffusion into equation (\ref{eq:H_d}) gives the millimeter-sized particles' scale height ($H_{\rm d}$).

\begin{figure*}
\begin{minipage}{17cm}
%\begin{figure}%[!ht]
\begin{center}
\includegraphics[width=8cm]{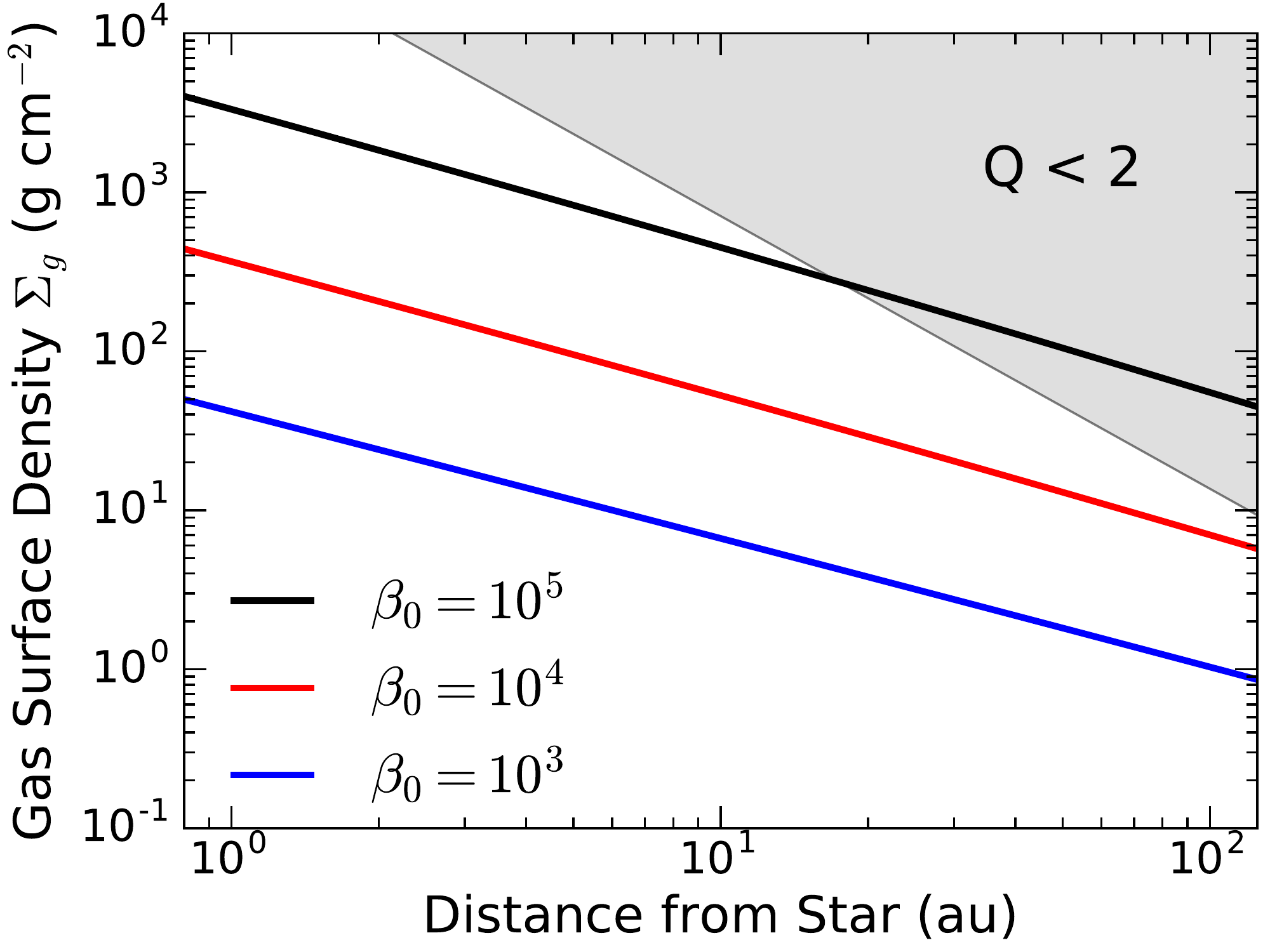}
\includegraphics[width=8cm]{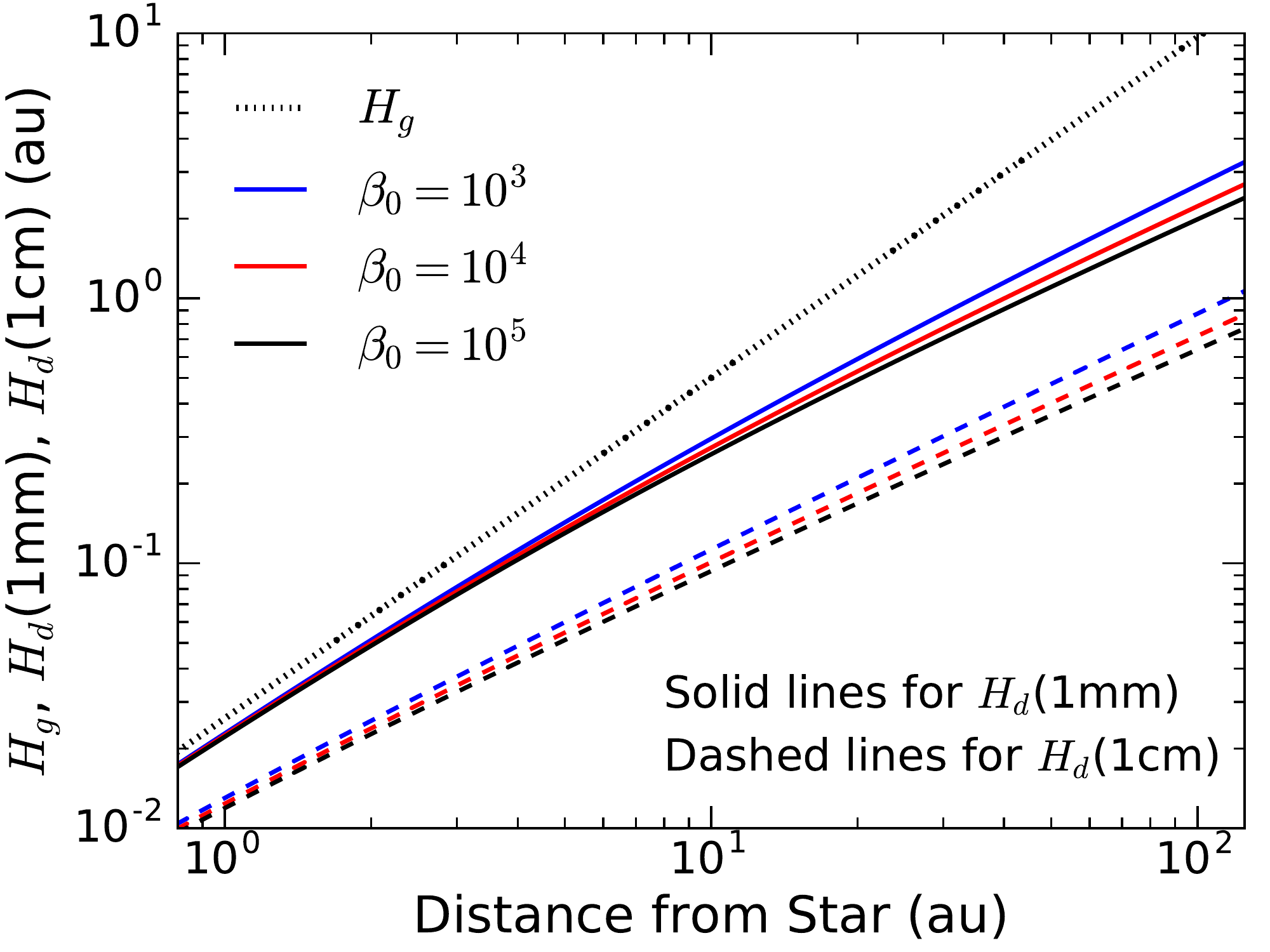}
\caption{Modeled gas and dust distributions for three strengths of the net vertical magnetic field, specified using the midplane plasma beta ($\beta_0$).
On the left, $\Sigma_{\rm g}$ is shown as a function of $r$. 
For comparison, the region where GI can operate is denoted by the shaded region.
The value of $\Sigma_{\rm g}$ increases with increasing $\beta_0$.
A relatively strong magnetic field ($\beta_0 \la 10^4$) is  needed to suppress GI in the disk.
On the right, the values of $H_{\rm g}$, $H_{\rm d}(1 \mbox{mm})$, and $H_{\rm d}(1 \mbox{cm})$ are shown
by the dotted, the solid, and the dashed lines, respectively.
The same-sized dust particles experience the roughly same degree of dust settling for a wide range of $\beta_0$.}
\label{fig2}
\end{center}
%\end{figure}
\end{minipage}
\end{figure*}

Figure \ref{fig2} shows the results.  
The left panel has gas surface density profiles over distance from the star for three values of $\beta_0$, 
while the right panel has the corresponding scale heights for gas and dust particles of millimeter and centimeter sizes.
In our formalism, self-gravity is neglected.
Accordingly, the model breaks down in the shaded region, 
where the Toomre parameter ($Q = c_{\rm s} \Omega /(\pi G \Sigma_{\rm g})$) is less than two, 
and self-gravity drives spiral waves or outright collapse \citep[e.g.,][]{vb07,sc14}.
Our results show that when the disk accretion rate is given, 
$\Sigma_{\rm g}$ becomes an increasing function of $\beta_0$.
This occurs because for a larger value of $\beta_0$, $\alpha_{\rm SS, eff}$ becomes smaller (see Figure \ref{fig1}).
In order to achieve a certain value of $\dot{M}$, $\Sigma_{\rm g}$ should increase (see equation (\ref{eq:mdot_b_s})).
As $\beta_0$ decreases (i.e.,  the case of stronger magnetic fields), disk winds can contribute to the disk angular momentum transport more.
This leads to a lower value of $\Sigma_{\rm g}$ even if the disk accretion rate is high.
We find that a relatively smaller value of $\beta_0 (\la 10^4)$ is needed to prevent gravitational instability (GI) from operating in the disk \citep[also see][]{lg15}.
For $H_{\rm d}$, 
our results show that both $H_{\rm d}(1 \mbox{mm})$ and $H_{\rm d}(1 \mbox{cm})$ are smaller than $H_{\rm g}$.
This is a simple reflection of dust settling, where turbulent mixing for the vertical direction is not significant enough 
for large-sized dust particles to catch up with the vertical distribution of gas (see equation (\ref{eq:H_d})).
We also find that the same-sized particles have the roughly same value of $H_{\rm d}$ even if $\beta_0$ varies.
This arises because the dependence on $\beta_0$ cancels out in equation (\ref{eq:H_d}) 
since ${\rm St} \propto \Sigma_{\rm g}^{-1} \propto \alpha_{\rm SS, eff} \propto W_{z\phi} \propto \beta_0^{-1}$, 
while $\alpha_D \propto \beta_0^{-1}$.

Thus, magnetically driven disk accretion can reproduce the high accretion rate observed for the HL Tau disk
when the disk is threaded by a vertical magnetic field with $\beta_0 \la 10^4$.
Furthermore, this disk model provides a natural explanation for the observed vertically thin layer of millimeter-sized dust particles.

\begin{figure*}
\begin{minipage}{17cm}
%\begin{figure}%[!ht]
\begin{center}
\includegraphics[width=8cm]{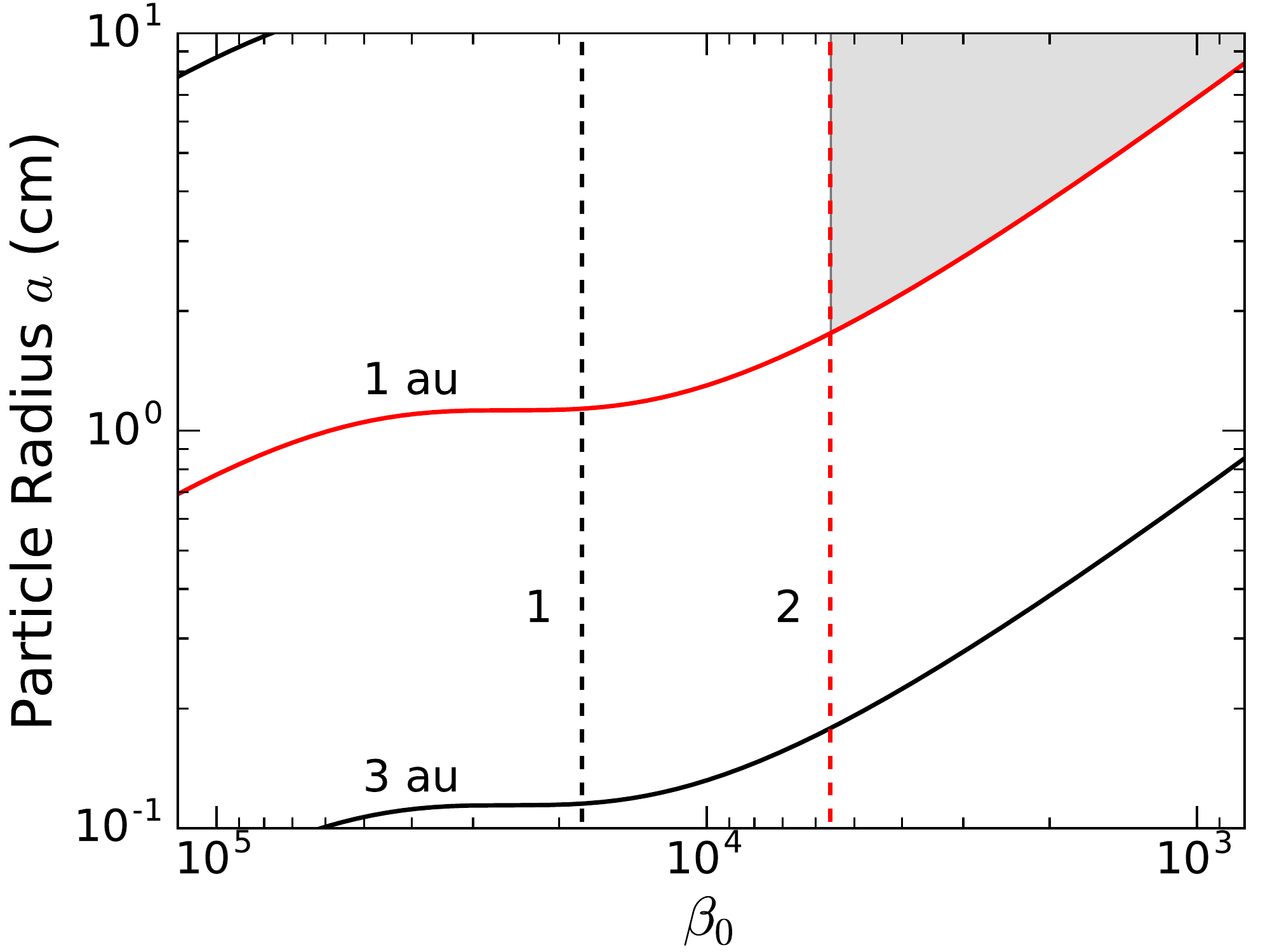}
\includegraphics[width=8cm]{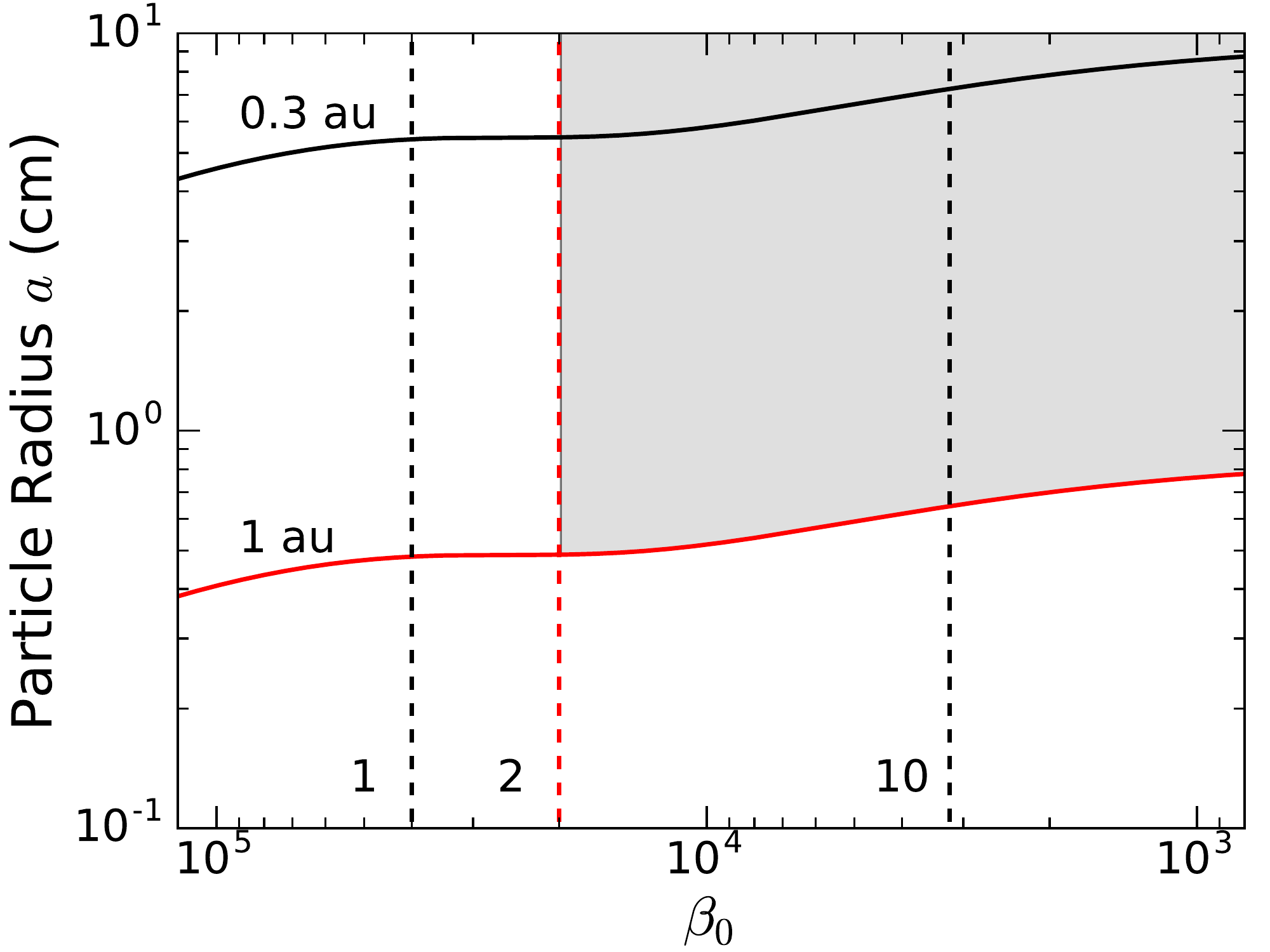}
\caption{Radius ($a$) of the particles whose scale height ($H_{\rm d}$) takes various values at r=100 au, as a function of $\beta_0$.  
Each scale height is a different solid line.  
Vertical dashed lines denote the values of $\beta_0$ at which the Toomre $Q$ parameter is 1, 2, and 10.  
For the left panel, the accretion stress comes from turbulence alone ($\alpha_{\rm SS}$, equation (\ref{eq:alpha_ss})),
while on the right, both turbulent and wind stresses are included ($\alpha_{\rm SS, eff}$, equation (\ref{eq:alpha_ss_eff})).
Because disk winds transport significant angular momentum when the magnetic fields are strong (i.e. when $\beta_0$ is smaller), 
the model in the right panel meets the constraints $H_{\rm d} \la 1$ au and $Q \ga 2$ over a bigger region of the parameter space.}
\label{fig3}
\end{center}
%\end{figure}
\end{minipage}
\end{figure*}

\subsection{Parameter study}

We now explore a broader parameter space, varying the particle size ($a$) and Stokes number ($\mbox{St}$) alongside the plasma beta.
In this parameter study, we compute both $a$ and $\mbox{St}$ that are needed to give a specified particle scale height ($H_{\rm d}$, see equation (\ref{eq:H_d})), 
as functions of $\beta_0$ at $r=100$ au.
These calculations are carried out to investigate how the above results (see Figure \ref{fig2}) can be maintained for a wide range of $\beta_0$.
In order to explicitly calibrate the importance of disk winds,
we here consider two cases: for the first case, $\alpha_{\rm SS}$ is used in computing $a$ and $\mbox{St}$ (see equation (\ref{eq:alpha_ss})),
and for the other case, $\alpha_{\rm SS,eff}$ is adopted (see equation (\ref{eq:alpha_ss_eff})).

Figure \ref{fig3} shows the results of $a$ with $\alpha_{\rm SS}$ and $\alpha_{\rm SS,eff}$ on the left and the right panels, respectively.
The solid lines denote the resulting values of $a$ for different dust heights ($H_{\rm d} =0.3$ au, 1 au, and 3 au at $r=100$ au),
and the vertical dashed lines are for the values of $\beta_0$ 
at which the corresponding Toomre $Q-$parameter becomes 1, 2, and 10 at $r=100$ au, respectively (also see Figure \ref{fig1}).
Our results show that a parameter space that satisfies both $H_{\rm d} \la 1$ au and $Q \ga 2$ at $r=100$ au can expand 
when disk winds are taken into account (see the shaded regions on both panels).
This is simply because only when MHD turbulence is considered, 
a lower value of $\beta_0$ is required to achieve that $\dot{M}=10^{-7}$ M$_{\odot}$ yr $^{-1}$.
This in turn pumps up the diffusion coefficient ($\alpha_D$, see Figure \ref{fig1}).
Then, a higher degree of dust settling can be established only for larger-sized ($a \ga 1$ cm) dust particles.
Consequently, the shaded region shrinks in terms of both $a$ and $\beta_0$ (see the left panel).
For the case that disk winds are treated properly in the accretion stress ($\alpha_{\rm SS,eff}$), however,
a modest value of $\beta_0$ is good enough to satisfy both a higher value of $\dot{M}$ and a lower value of $H_{\rm d}$.
This is attributed to that disk winds can remove the disk angular momentum considerably.
As a result, when $\alpha_{\rm SS, eff}$ is adopted, 
a parameter space becomes bigger, where both $H_{\rm d} \la 1$ au and $Q \ga 2$ at $r=100$ au are fulfilled simultaneously.  

\begin{figure*}
\begin{minipage}{17cm}
%\begin{figure}%[!ht]
\begin{center}
\includegraphics[width=8cm]{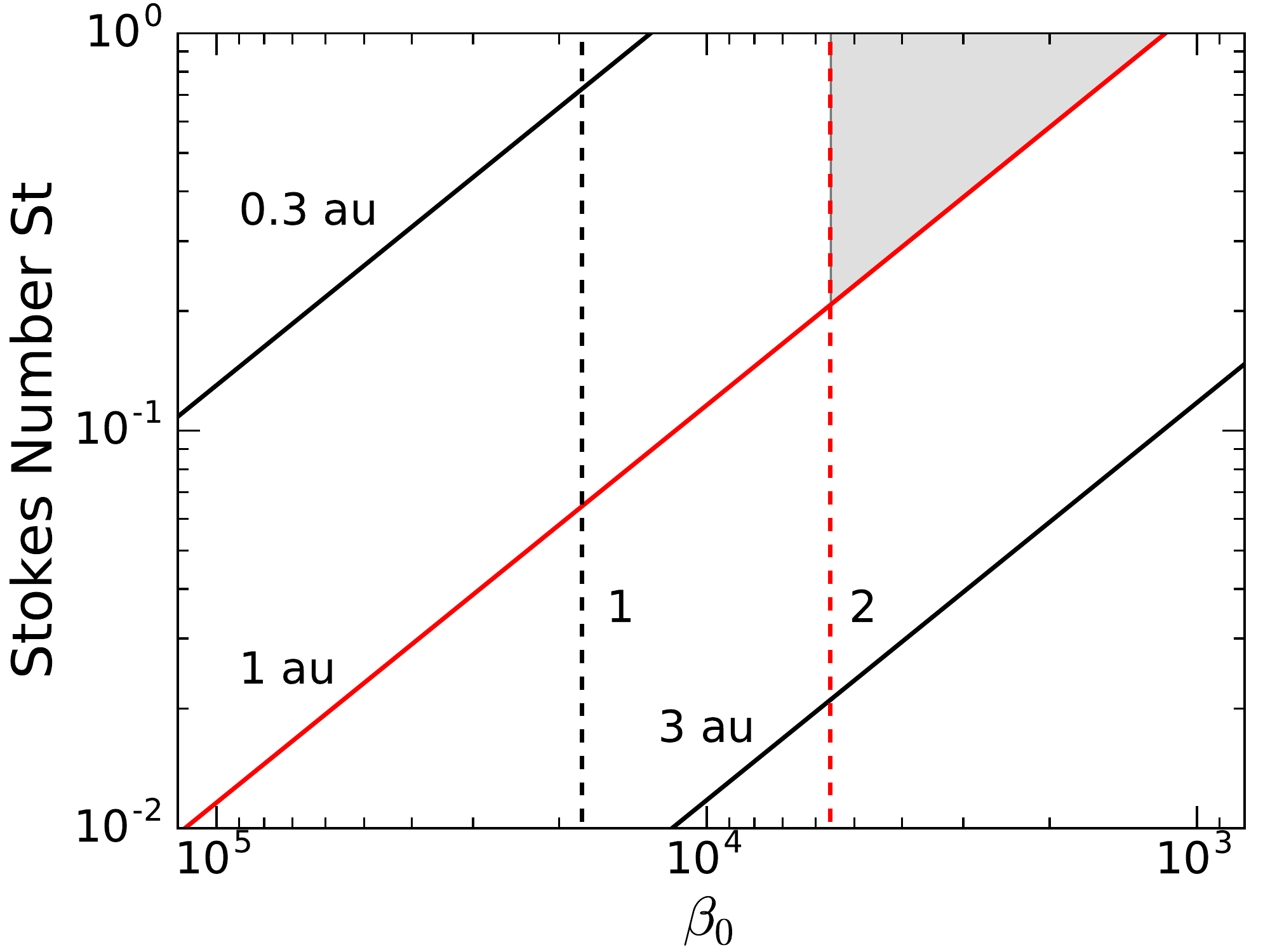}
\includegraphics[width=8cm]{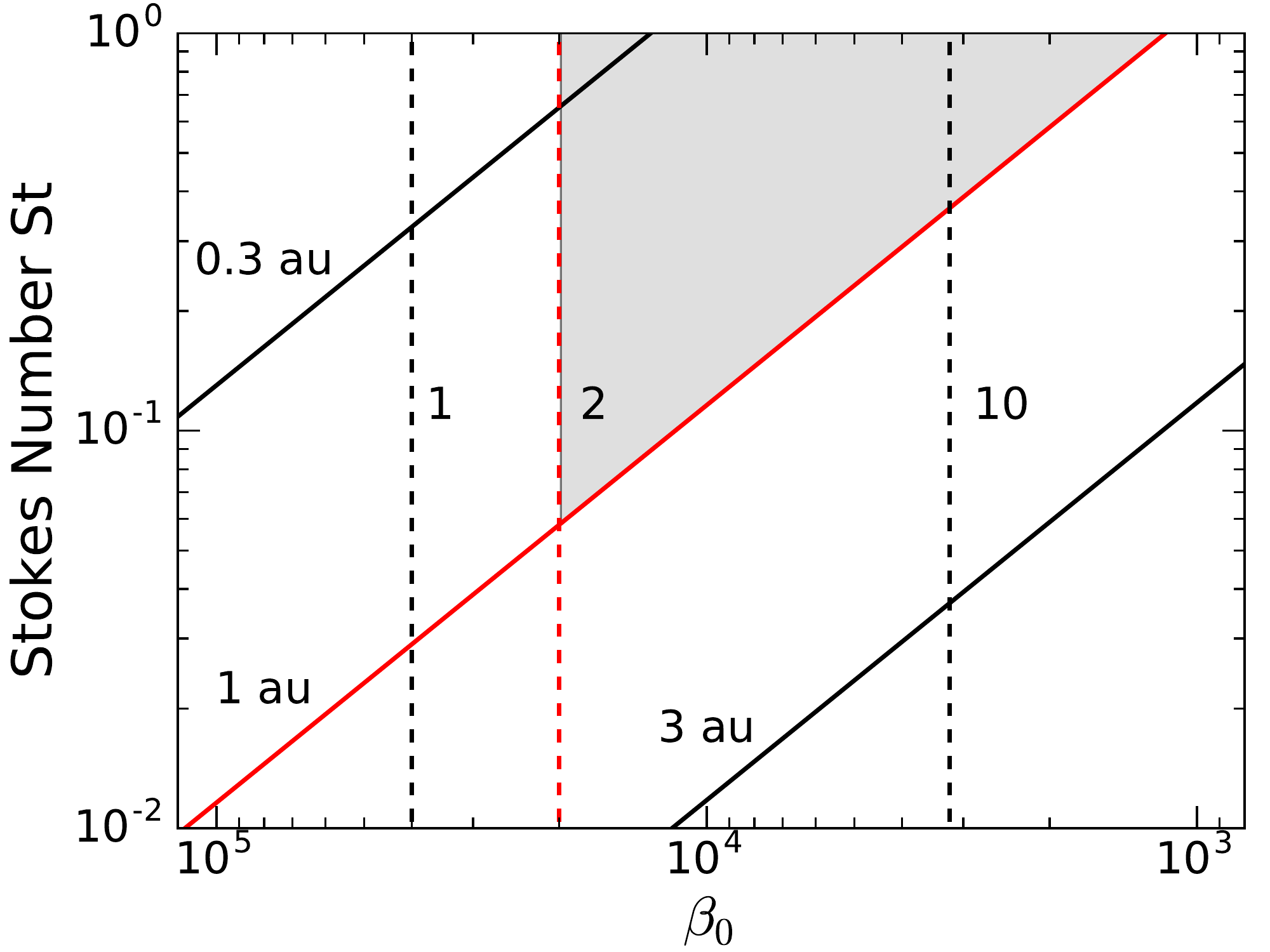}
\caption{Stokes number ($\mbox{St}$) of the particles whose scale height ($H_{\rm d}$) takes various values at 100 au, as a function of $\beta_0$.  
As in Figure \ref{fig3}, each scale height is a different solid line, and vertical dashed lines denote $\beta_0$ for which the Toomre $Q$ parameter is 1, 2, and 10.  
Turbulence is only taken into account on the left panel ($\alpha_{\rm SS}$, equation (\ref{eq:alpha_ss})),
and both turbulent and disk winds are considered on the right panel ($\alpha_{\rm SS, eff}$, equation (\ref{eq:alpha_ss_eff})).
The model is valid and agrees with the observed particle layer thickness if $H_{\rm d} \la 1$ au and $Q \ga 2$.  
Because radial drift governs the particles' growth and movements, the Stokes number also should be near 0.1.  
These three conditions are met simultaneously only when disk winds transport significant angular momentum (see the right panel), 
and the disk is initially threaded by a relatively strong magnetic field ($\beta_0 \approx 2 \times 10^4$).}
\label{fig4}
\end{center}
%\end{figure}
\end{minipage}
\end{figure*}

While Figure \ref{fig3} shows that $a$ is relatively insensitive to the change of $\beta_0$ in the shaded region for the case of $\alpha_{\rm SS,eff}$,
it is interesting to examine whether a much tighter constraint can be obtained on the value of $\beta_0$.
For this purpose, we also plot the resulting value of $\mbox{St}$ as a function of $\beta_0$ (see Figure \ref{fig4}).
This is motivated by the calculations of dust growth which show that
when dust growth is limited by radial drift,
the maximum size of dust particles is peaked around $\mbox{St} \simeq 0.1$ \citep[e.g.,][]{bdh08,bdb10,otk12}.
Since radial drift is one of the severest barriers for dust growth in protoplanetary disks,
it is important to verify whether or not our results can satisfy the condition that $\mbox{St} \simeq 0.1$,
which may serve as an additional constraint.

\begin{figure}%[!ht]
\begin{center}
\includegraphics[width=8cm]{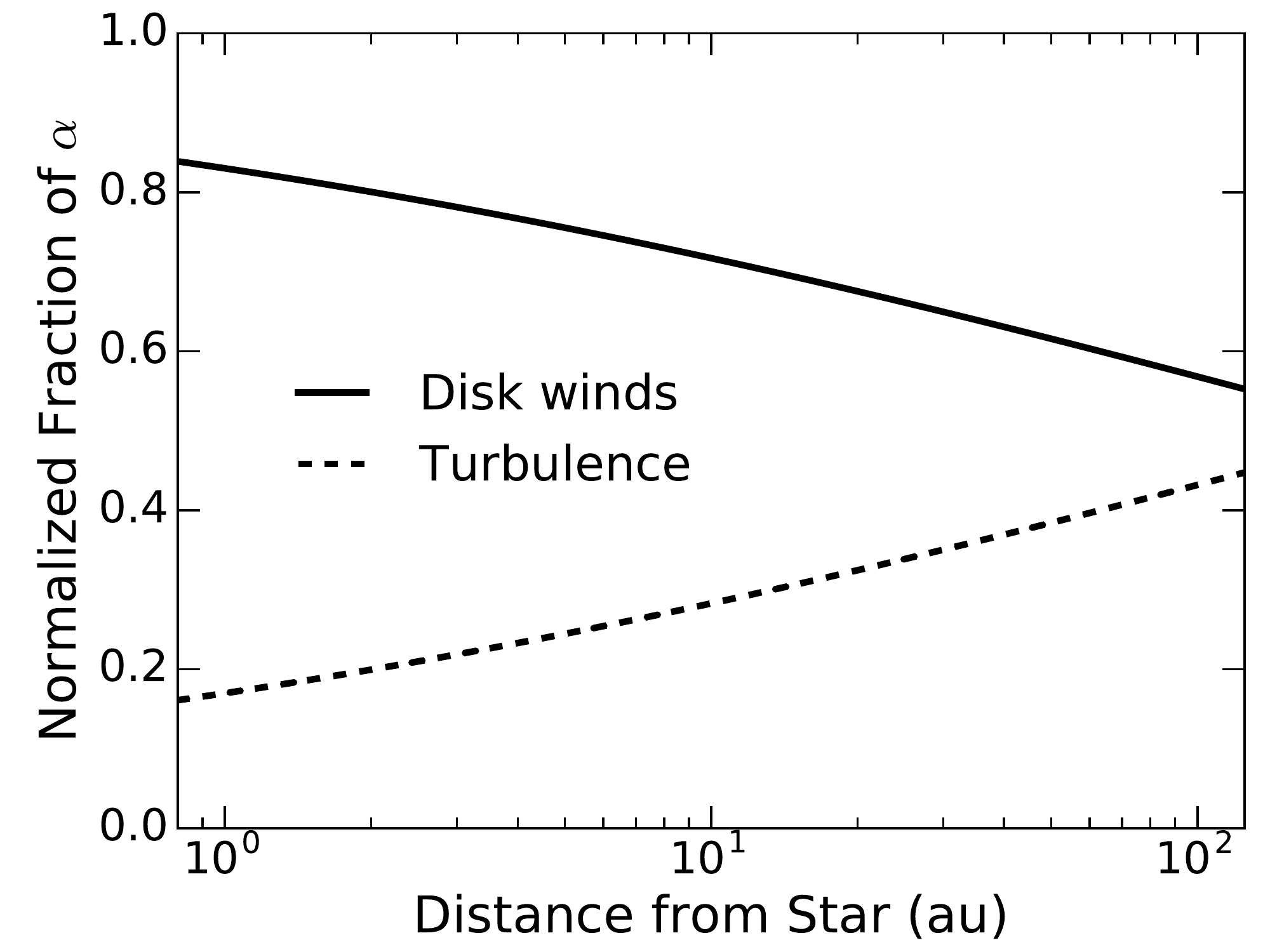}
\caption{The contribution of the turbulent ($\alpha_{\rm SS} / \alpha_{\rm SS, eff}$) and the wind ($1-\alpha_{\rm SS} / \alpha_{\rm SS, eff}$) stresses
to $\alpha_{\rm SS, eff}$ as a function of the distance from the central star 
for the best case ($\beta_0 = 2 \times 10^4$, see the right panel of Figure \ref{fig4}).
The wind stress dominates over the turbulent one at the entire region of the disk when disk winds are taken into account properly (equation (\ref{eq:alpha_ss_eff})).
The contribution from the wind stress decreases as increasing the distance from the central star.
This is because the factor ($r / H_{\rm g}$) becomes smaller at larger disk radii for the so-called flared disk model.}
\label{fig5}
\end{center}
\end{figure}

Figure \ref{fig4} shows the results.
We first point out that the behavior of $\mbox{St}$ is the same with and without disk winds.
This is simply because $\mbox{St}$ is s function of only $\beta_0$ when $H_{\rm d}$ and $H_{\rm g}$ are given (equation (\ref{eq:H_d})).
We then see that all three conditions ($H_{\rm d} \la 1$ au and $Q \ga 2$ at $r=100$, and $\mbox{St} \simeq 0.1$) 
are realized simultaneously, when disk winds are properly included (right panel).
We therefore suggest that disk winds are necessary to account for the configuration of the HL Tau disk.
In fact, we find that the wind stress contributes to the value of $\alpha_{\rm SS, eff}$ by more than 50 \% 
across the entire region of the disk in our model (see Figure \ref{fig5}).
It is, however, important to emphasize that only a small region of parameter space meets all the conditions 
even for the case of $\alpha_{\rm SS,eff}$ (see Figure \ref{fig4}).

%\begin{figure*}
%\begin{minipage}{17cm}
%\end{minipage}
%\end{figure*}

Thus, our results suggest that in order to reproduce the disk configuration observed toward HL Tau,
the presence of disk winds is inevitable and the initial vertical magnetic field should be relatively strong ($\beta_0 \simeq 2 \times 10^4$).

\subsection{The global structure of vertical magnetic fields}

Based on the above results, we obtain the best value of $\beta_0$.
In the following, we will make use of the value to investigate the resulting disk properties such as vertical magnetic fields and the gas-to-dust ratio.
In this section, we compute the global structure of the initial, vertical magnetic field ($B_{z,0}$).
This becomes possible, because $B_{z,0}$ can be derived from the resulting $\Sigma_{\rm g}$ and $\beta_0$ for a given value of $c_{\rm s}$ (see equation (\ref{eq:beta})).

\begin{figure}%[!ht]
\begin{center}
\includegraphics[width=8cm]{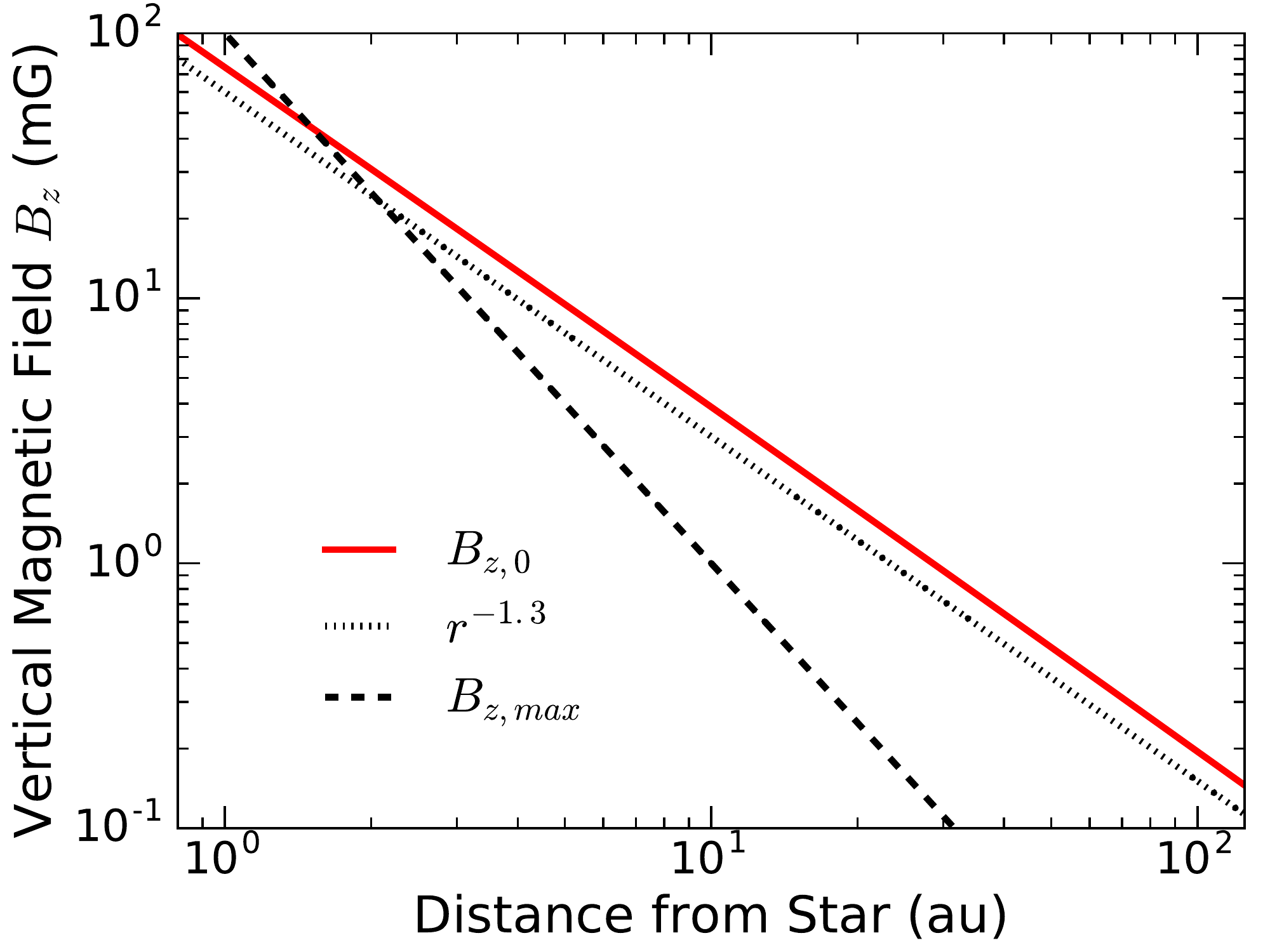}
\caption{The resulting, initial vertical magnetic field ($B_{z,0}$) as a function of the distance from the central star for the best case ($\beta_0= 2 \times 10^4$).
Our results show that the profile (the solid line) is characterized well by $r^{-1.3}$ (the dotted line), 
suggesting that the disk accretion rate is determined mainly by the value of $B_{z,0}$
and that the $\Sigma$ dependence cancels out in $\dot{M}$.
The comparison between our profile with the previous study ($B_{z, \rm max}$, the dashed line) 
implies that the vertical magnetic flux in the outer part of the HL Tau disk may be higher than that in protoplanetary disks
either due to the presence of the envelope or due to the intrinsic difference between the star formation and the planet formation stages.}
\label{fig6}
\end{center}
\end{figure}

Figure \ref{fig6} shows the profile of $B_{z,0}$ for our best case ($\beta_{0} = 2 \times 10^4$).
We find that the profile is characterized by $B_{z,0} \propto r^{-1.3}$,
and can be reproduced by the following calculation:
as discussed above (see Figure \ref{fig5}), disk winds play the dominant role in transporting the disk angular momentum.
Then, it can be assumed that $\alpha_{\rm SS, eff} \sim (r/H_{\rm g}) W_{z\phi} \sim (r/H_{\rm g}) \beta_{0}^{-1}$ (see equations (\ref{eq:alpha_ss_eff}) and (\ref{eq:W_zphi})).
As a result, the disk accretion rate can be given as (see equation (\ref{eq:mdot_b_s}))
\begin{equation}
\label{eq:mdot_approx}
\dot{M} \sim \frac{\Sigma_{\rm g} c_{\rm s}^2}{\Omega} \frac{r}{H_{\rm g} \beta_0} \sim \frac{r}{\Omega} B_{z,0}^2.
\end{equation}
Since the steady disk accretion is currently considered ($\dot{M} = \mbox{const.}$), the resulting magnetic field profile becomes $B_{z,0} \propto r^{-1.25}$.

This derivation is interesting in the sense that 
the $\Sigma_{\rm g}$ dependence cancels out when computing the disk accretion rate (see equation (\ref{eq:mdot_approx})).
In other words, the disk accretion rate is determined mostly by the initial $B_{z,0}$.
This finding is consistent with the previous results of ideal MHD simulations \citep{hgb95}, and of non-ideal MHD simulations \citep{oh11,b13}.
Furthermore, our profile itself is comparable with the previous studies \citep{otm14};
for instance, \citet{oh11} perform resistive MHD simulations and find that $B_{z,0} \propto r^{-1.38}$
while \citet{b13} perform resistive MHD simulations with ambipolar diffusion and find that $B_{z,0} \propto r^{-1.44}$ \citep[see][for the correct formula]{b13_c}.

The absolute magnitude of $B_{z,0}$ is more uncertain.
This is because the value of $B_{z,0}$ should be controlled by 
the balance between the radial dragging and diffusion of magnetic fields \citep[e.g.,][]{lpp94,g014,otm14,to14}.
Here, we compare the above results with those obtained by \citet{otm14}.
In their study, the upper limit of large scale magnetic fields ($B_{z,\rm max}$) are derived by considering the steady state,
where radial inward advection of magnetic fields balances out with their outward diffusion.
Mathematically, $B_{z,\rm max}$ can be given as (see their equation (48))
\begin{equation}
 B_{z, \rm max} = 100 \left( \frac{r}{\mbox{au}} \right)^{-2} \left( \frac{r_{\rm out}}{100 \mbox{ au}} \right)^2  \left( \frac{B_{\infty}}{ 10 \mu \mbox{G}} \right) \mbox{mG},
\end{equation}
where $B_{\infty}$ is the vertical magnetic field at $r=r_{\rm out}$.
Figure \ref{fig6} summarizes our comparison. We find that our $B_{z,0}$ is much larger than the value of $B_{z,\rm max}$ at the outer part of the disk.
Since the absolute magnitude of $B_{z,\rm max}$ is scaled by the magnetic field strength at the outer edge of disks,
this difference may come from either that the presence of the envelope may trigger more efficient, inward advection of magnetic fields
or that disks in the star formation stage may intrinsically possess a more strong field than those in the planet formation stage.
Note that \citet{frd15} perform global, non-ideal MHD simulations to model the detailed structure of protoplanetery disks. 
In these simulations, they adopt the initial $B_z$ that is similar to our $B_{z,0}$, and find that some observable structures such as gaps and rings are generated.
It is obvious that more detailed calculations would be needed to fully address this point.

Thus, a coupling of our model with the ALMA observations of HL Tau has made it possible to discuss the global structure of magnetic fields threading disks 
in a self-consistent fashion.

%\begin{figure*}
%\begin{minipage}{17cm}
%\end{minipage}
%\end{figure*}

\subsection{The gas and dust distributions in the outer part of the disk}

In this section, we consider the gas and dust distributions in the HL Tau disk.

To proceed, we compare our results with those of \citet{klm11}.
As already described in Section \ref{sec:model_kwon},
the surface density distribution of dust ($\Sigma_{\rm d}$) for the HL Tau disk can be represented well by similarity solutions,
except for the gap regions \citep[see equation (\ref{eq:sigma_d}),][]{klm15,ahh16,pdm16}.
It should be noted that the dust thermal emissions observed by both ALMA and CARMA are optically thin only in the outer part of the HL Tau disk \citep{bph15,chc16}.
Accordingly, we here focus on the surface density profile at $r > 40$ au.

%\begin{figure*}
%\begin{minipage}{17cm}
\begin{figure}%[!ht]
\begin{center}
\includegraphics[width=8cm]{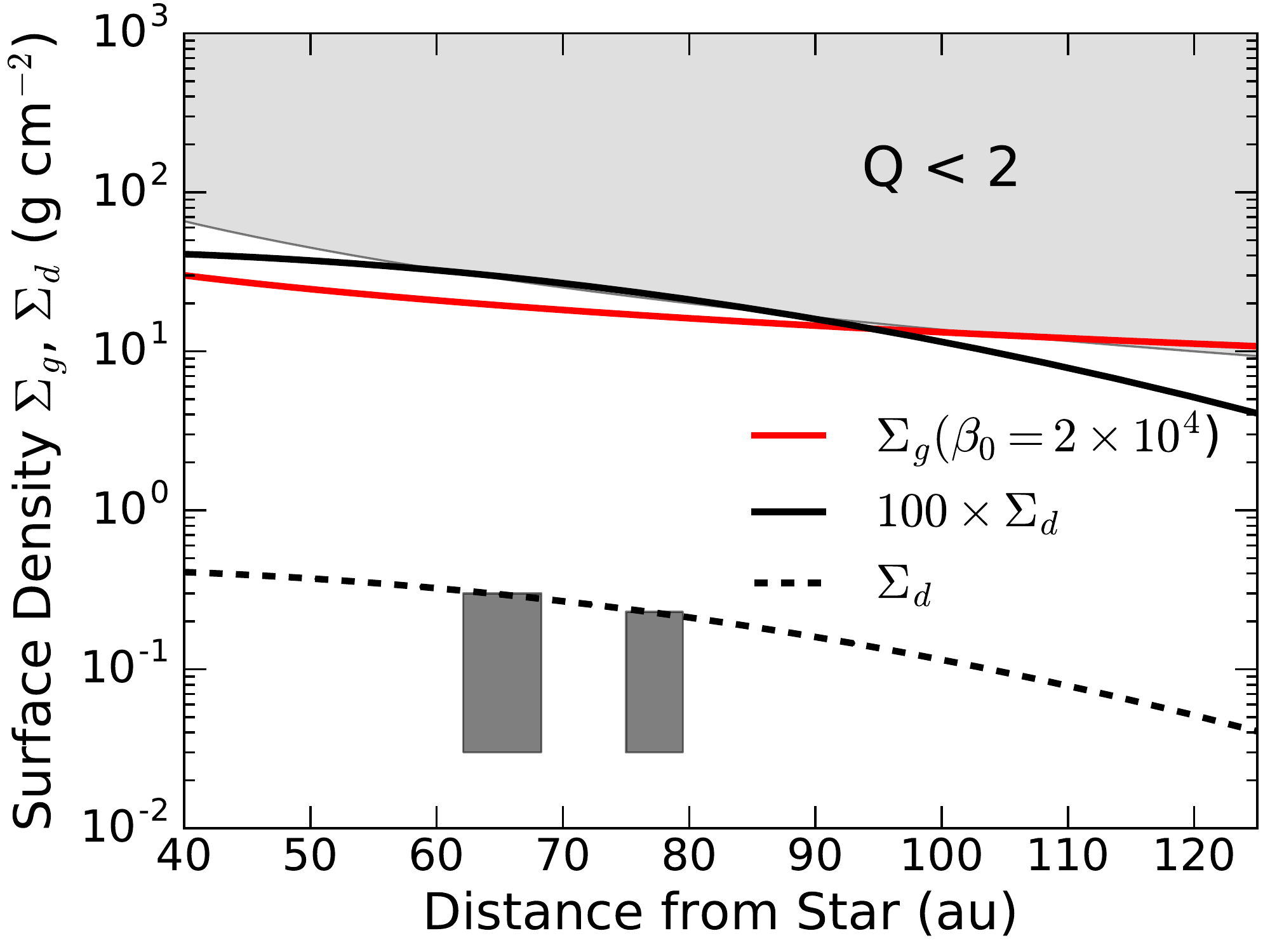}
\caption{The model's surface density profile compared with other estimates. 
Our best-matching case having $\beta_0=2\times 10^4$ (red curve) is plotted 
alongside the dust profile from CARMA millimeter interferometric measurements \citep[black dashed]{klm11}
which is also scaled up by a factor 100 (black solid) to compare with the gas profile.  
Dark shaded bands mark the dust gaps observed with ALMA.  
The depth of these gaps is adopted from the model results of \citet{pdm16}.
The red and black solid curves' differing slopes suggest the gas-to-dust ratio could be lower inside 100 au than outside, 
as expected if the millimeter-sized particles either undergo radial drift, or form later here owing to the long dynamical timescales.}
\label{fig7}
\end{center}
\end{figure}
%\end{minipage}
%\end{figure*}

Figure \ref{fig7} shows our results for the best case ($\beta_0 =2 \times 10^4$, the red solid line), 
$10^2 \times \Sigma_{\rm d}$ (the black, solid line), and $\Sigma_{\rm d}$ (the dashed line).
The dust gaps observed by ALMA are also denoted by the dark, shaded regions.
Notice that if the gas-to-dust ratio is 100, the Kwon et al. measurements imply marginal gravitational instability at 60-80 au around the gaps (see $10^2 \times \Sigma_{\rm d}$).
Our model avoids GI because of a slightly lower surface density.
In other words, the gas-to-dust ratio may be lower than 100 there.
On the contrary, our model suggests that the gas-to-dust ratio may be higher than 100 beyond $r =100$ au
since $\Sigma_{\rm g} > 10^2 \times \Sigma_{\rm d}$.
Our results therefore imply that the gas-to-dust ratio may increase from the intermediate region of disks to the outer one.
It is interesting that this implication is indeed consistent with the observational trend of protoplanetary disks:
larger dust particles tend to concentrate toward the central star there \citep[e.g.,][]{wc11}.
Based on detailed radiative transfer modeling, \citet{pdm16} have also suggested that 
larger dust particles may be depleted at $r\simeq100$ au in the HL Tau disk
possibly due to either (efficient) radial drift of such particles or slower growth of them.

Thus, the results of our modeling are useful to provide some implications for the gas-to-dust ratio in the HL Tau disk.

\section{Discussion} \label{sec:disc}

We have so far focused on two of the key properties of the HL Tau disk: 
the high disk accretion rate ($\sim 10^{-7}$ M$_{\odot}$ yr$^{-1}$) onto the central star and the vertical thin ($\sim 1$ au) dusty layer at $r=100$ au.
We have demonstrated above that magnetically driven disk accretion can reproduce these features naturally
when disk winds are properly taken into account and the initial plasma $\beta$ is $\beta_0 \simeq 2 \times 10^4$.
It is however important that a number of other interesting features have been reported for the HL Tau disk.
Here we examine how our results are consistent with these features.
In addition, we discuss limitations of our model and other possible mechanisms of disk evolution for the HL Tau system.

\subsection{Other observational features}

In this section, we consider the observations of winds/outflow from the HL Tau disk, a polarization map for the disk, 
and the molecular line emission, and discuss how our disk model can account for these observational results.

It is well recognized that the HL Tau system exhibits large-scale ($\sim 10^4$ au), bipolar outflows \citep{mpl96}.
Observations trace the inner jet down to $\sim 100$ au from the disk \citep[e.g.,][]{phk06,bbm10} and 
the modest spatial resolution ($\sim 30$ au) ALMA observations are examined for characterizing the outward motion of molecular gas \citep[e.g.,][]{kmmj16}.
Despite such efforts, identification of the launching points of disk winds/outflow is still not done yet for the HL Tau disk observationally.
Recently, \citet{bwh16} have conducted ALMA observations toward TMC1A with the high spatial resolution ($\sim 6$ au)
and inferred that molecular ($^{12}$CO) outflows are very likely to originate from the Keplerian disk for this target.
Interferometric observations with similar resolutions are demanded for the HL Tau system to specify the footpoint radius of outflows
and to verify the importance of disk winds.

Particles aligned in a magnetic field can produce polarized thermal continuum emission at sub-millimeter wavelengths \citep[e.g.,][]{bw17}.
Mapping polarization vectors across the HL Tau disk with modest spatial resolution of about 80 au,
\citet{slk14} inferred that the field's strongest component is toroidal.  
A strong toroidal field was also found in synthetic dust thermal polarization maps that are generated from global 3D MHD simulations \citep{bfw17}.
On the contrary, \citet{mtk16} found that MHD simulations with a large radial component better reproduced the observed polarization map for the HL Tau disk.
The presence of magnetic fields for the radial direction would be viewed as one of the evidence that magnetically induced disk winds are launched.
Therefore, the polarization observation suggests that disk winds are very likely to play an important role in the HL Tau disk.
Note that their simulations are carried out under the ideal MHD limit and adopt a very low value of $\beta_0(\simeq 2.5)$.
On the other hand, our results are based on the non-ideal MHD simulations, and find that $\beta_0 \sim 10^4$ for the base case.
We thus cannot compare our results with those of \citet{mtk16} directly.
It is nonetheless interesting that both studies lead to the conclusion that disk winds would be crucial for the HL Tau disk.

Furthermore, \citet{mtk16} have provided the following implication: 
given that the radial component becomes important only in the disk surface,
small ($\la$ 0.1 mm) grains that are located in the upper layer of disks should contribute significantly to the observed polarized emission. 
This means that the maximum size of dust particles in the disk midplane may be too large to contribute to the polarized emission.
Taking into account that the polarized emission at (sub)mm wavelengths  can be generated by dust particles with the size of $\la 0.1$ mm \citep[e.g.,][]{cl07},
our results become compatible with their implication; the majority of dust in the midplane may be larger than 1 mm in size (see Figure \ref{fig3}).
It should be pointed out that self-scattering of dust particles can also provide a reasonable explanation for the observed polarization map
without the presence of magnetic fields and dust alignment \citep{kmm16,yll16,tln17}.

We have so far discussed the observational features that can be used as a probe to detect magnetically induced disk winds.
This is because disk winds are needed as an alternative process to transport angular momentum 
when MHD turbulence is suppressed in the outer part of the HL Tau disk.
It would also be useful to directly infer the level of turbulence there, using other observables.
Observations of the molecular line emission can play such a role. 
In general, it is very difficult to provide some meaningful constraints on the level of disk turbulence 
by detecting the molecular line emission.
This is attributed from a need of the spatially and spectrally resolved measurements to calibrate the width of non-thermal broadening,
which can be less than a few percents of the width caused by the thermal broadening.
The recent advance in interferometric observations in (sub)millimeter wavelengths 
has enabled to conduct such observational studies \citep[e.g.,][]{hwa11,gdw12,fhr15}.
For instance, \citet{fhr17} use the ALMA data taken toward HD 163296, 
and derive stringent upper limits on the non-thermal motion for the disk.
They infer that the line-widths generated by turbulence is $\sim 5$ \% of the local sound speed.
It is interesting that when MRI and the resulting MHD turbulence fully operate in the disk,
these limits cannot be reproduced.
In other words, a level of turbulence should be weak in the outer part of the disk,
which may be caused by non-ideal MHD effects.
It is, however, important to recognize that the quality of the data is still not be high enough 
to fully address the strength of disk turbulence \citep[e.g.,][]{tgs16}.
In fact, the upper limit derived by \citet{fhr17} is still looser than the prediction of \citet{pdm16} that 
the velocity arising from turbulence is $\sim 1$ \% of the sound speed.
More observations are demanded to tightly infer the level of turbulence for protoplanetary disks including the HL Tau system.

Thus, while the spatially and spectrally resolved observational data are needed 
to reliably examine the validity of magnetically driven disk accretion models for the HL Tau system,
qualitative assessments suggest that a number of important observational features can be explained by our model.

\subsection{Limitations of our model} \label{disc_limit}

As discussed above, the observed properties of the HL Tau disk can be better understood by magnetically driven disk accretion models.
We must admit, however, that our disk model is developed under a number of assumptions.
Here, we discuss how these assumptions can affect our finding and what would be limitations of our model.

First, our model heavily relies on the results of non-ideal MHD simulations.
Specifically, the spatially integrated, temporally averaged values of $W_{r\phi}$, $W_{z\phi}$, and $\alpha_D$ are utilized 
to obtain the fitting formulae (see Figure \ref{fig1}).
This approach can end up with both overestimate and underestimate of $H_{\rm d}$ and $\Sigma_{\rm g}$,
depending on the spatial integration.
For instance, the midplane value of $\alpha_D$ can be overestimated by integrating it out up to a large value of the vertical height,
which can puff up dusty layers for the vertical direction.
On the other hand, the values of $W_{r\phi}$ and $W_{z\phi}$ at disk surfaces can be underestimated by swearing them out over the intermediate height,
which can lead to a higher value of $\Sigma_{\rm g}$.
Numerical simulations are therefore needed to fully confirm the results of our semi-analytical model.

Second, our model does not include the mass loss from disk surfaces that can be triggered by disk winds.
It is interesting that both numerical simulations and observations suggest that some degree of the mass loss arises from circumstellar disks.
In fact, \citet{kmmj16} use the ALMA data of CO emission line and 
estimate that the mass loss rate caused by outflow from HL Tau can be as high as the disk accretion rate.
Theoretically, both ideal and non-ideal MHD simulations confirm the mass loss due to disk winds \citep[e.g.,][]{si09,bs13,sba13}.
Since the mass loss rate can be computed as the product of the gas density and sound speed at the launching point,
the resulting value is quite sensitive to the location of the launching point.
In other words, it is still hard to derive a reliable value of the mass loss rate both observationally and theoretically.
Nonetheless, our models are very likely to overestimate the resulting value of $\Sigma_{\rm g}$ due to neglect of the mass loss.

\begin{figure*}
\begin{minipage}{17cm}
%\begin{figure}%[!ht]
\begin{center}
\includegraphics[width=8cm]{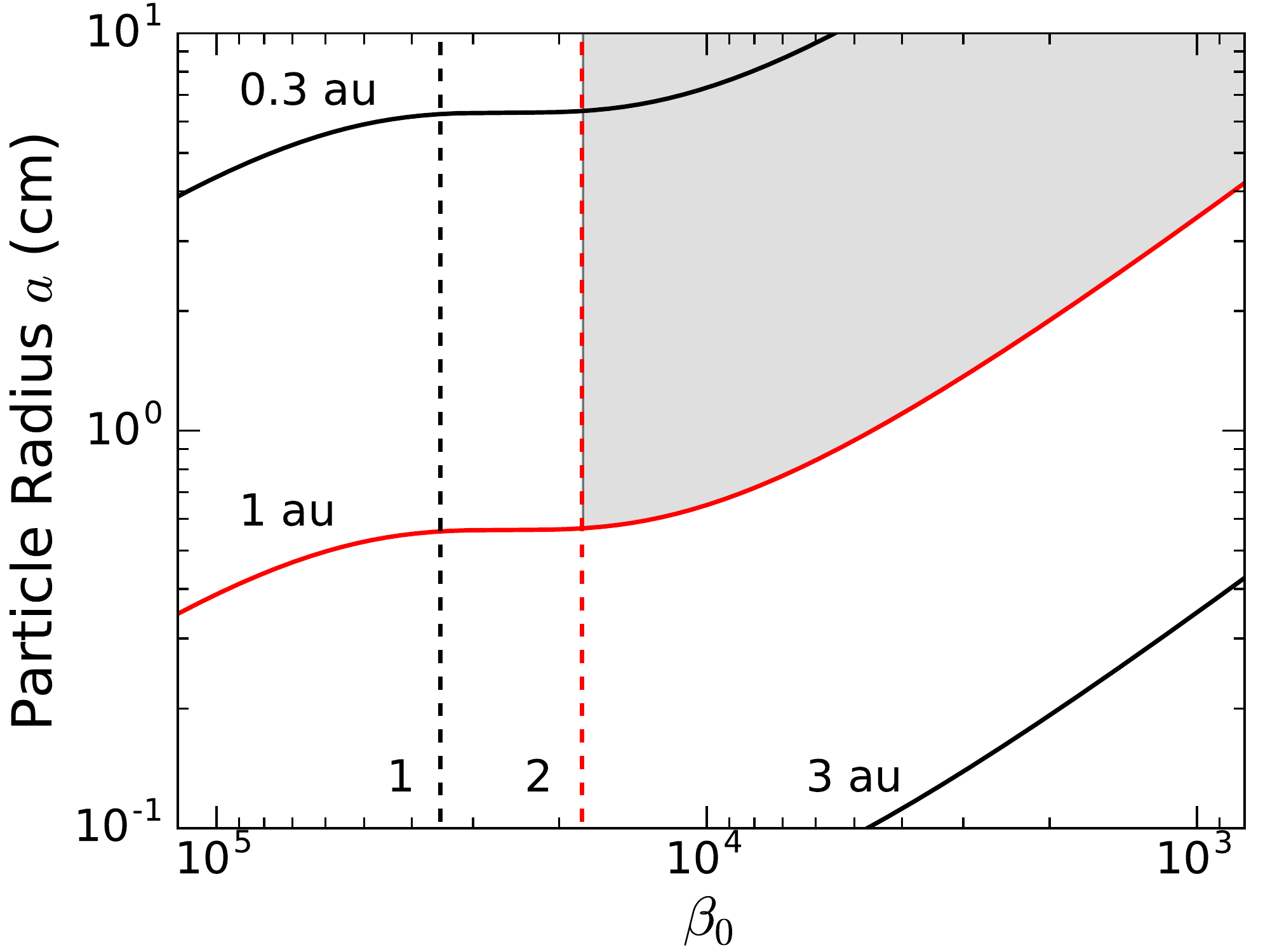}
\includegraphics[width=8cm]{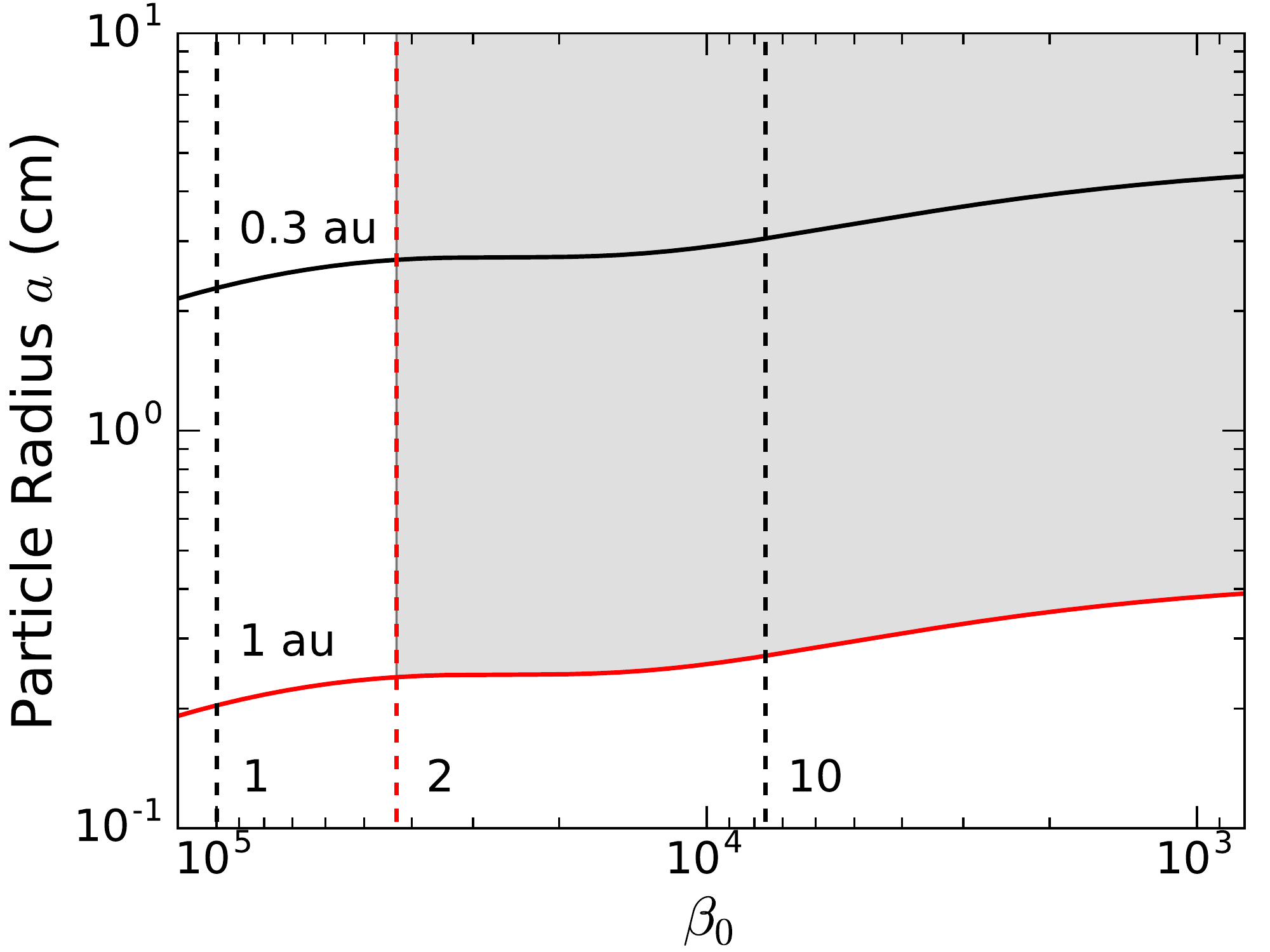}
\caption{Radius ($a$) of the particles whose scale height ($H_{\rm d}$) takes various values at r=100 au, as a function of $\beta_0$ 
(as in Figure \ref{fig3}).
The left panel shows the results for the case of turbulence ($\alpha_{\rm SS}$),
while the right panel is for the case of both turbulence and disk winds ($\alpha_{\rm SS, eff}$).
In these plots, it is assumed that $\dot{M} = 5 \times 10^{-8}$ M$_{\odot}$ yr$^{-1}$.}
\label{fig8}
\end{center}
%\end{figure}
\end{minipage}
\end{figure*}

\begin{figure*}
\begin{minipage}{17cm}
%\begin{figure}%[!ht]
\begin{center}
\includegraphics[width=8cm]{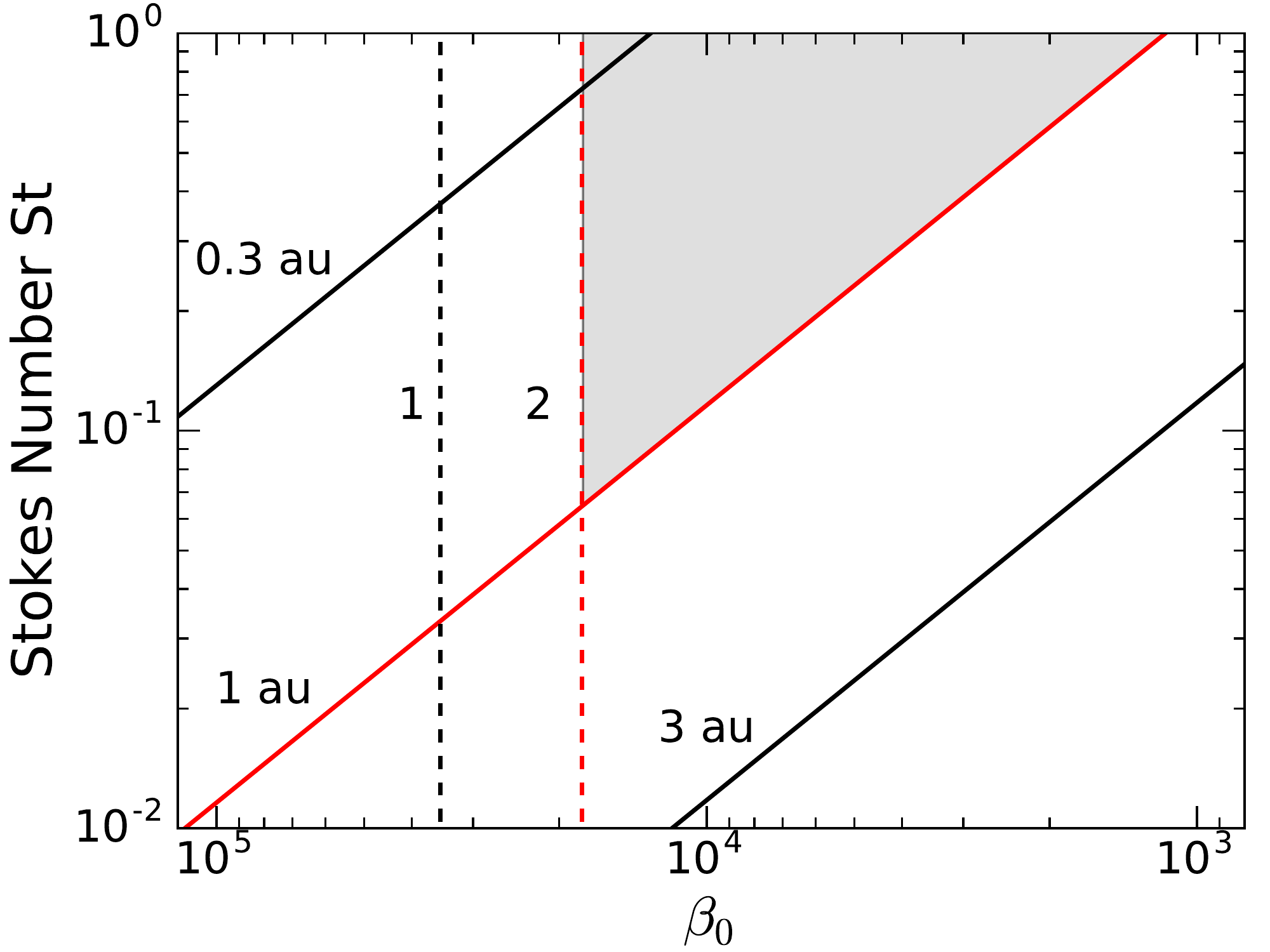}
\includegraphics[width=8cm]{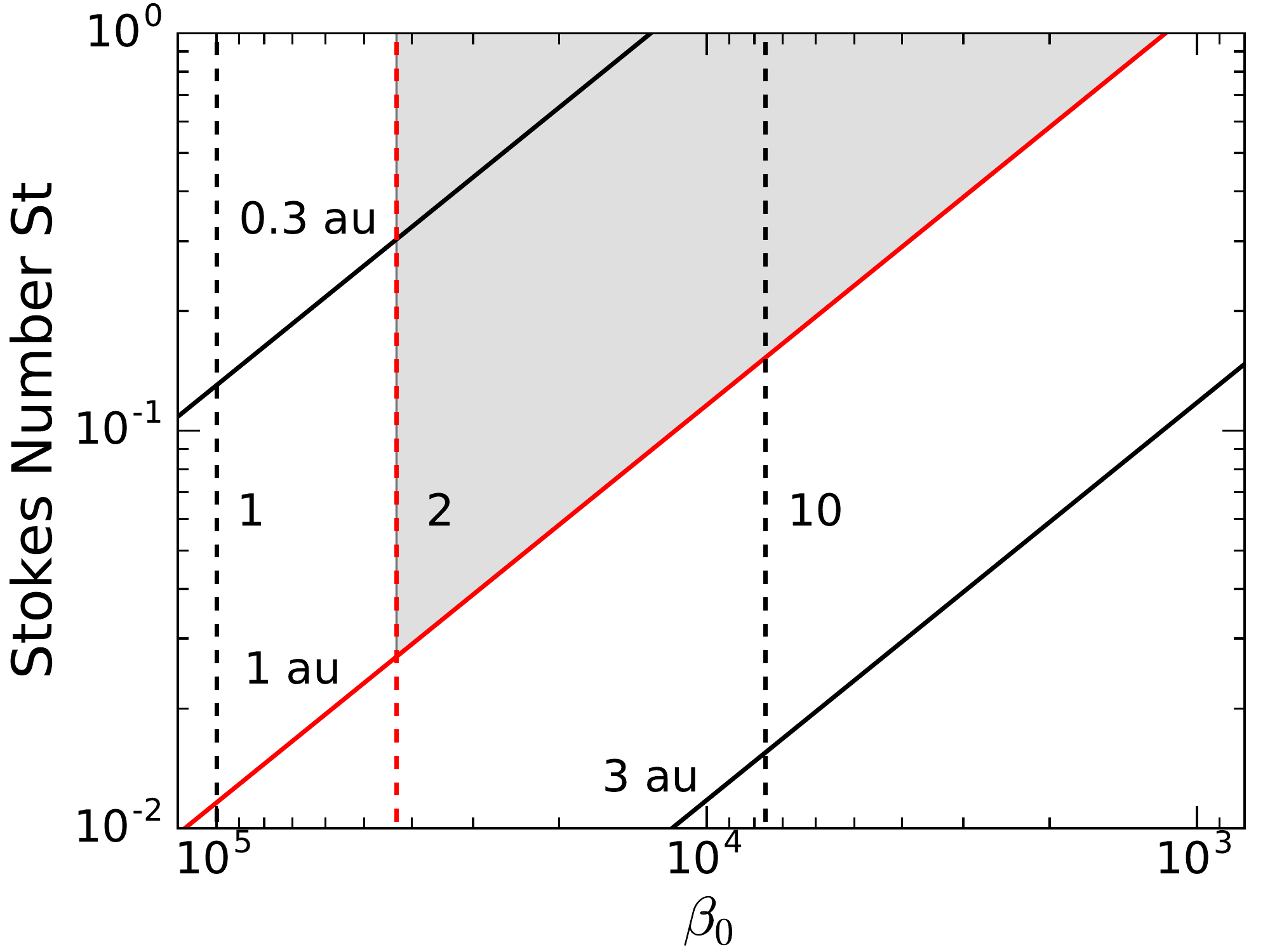}
\caption{Stokes number ($\mbox{St}$) of the particles whose scale height ($H_{\rm d}$) takes various values at 100 au, as a function of $\beta_0$ 
(as in Figure \ref{fig4}). 
The turbulence case ($\alpha_{\rm SS}$) is shown on the left panel,
and the turbulence plus disk wind case is on the right panel ($\alpha_{\rm SS, eff}$).
In these plots, it is assumed that $\dot{M} = 5 \times 10^{-8}$ M$_{\odot}$ yr$^{-1}$.}
\label{fig9}
\end{center}
%\end{figure}
\end{minipage}
\end{figure*}

Third, we have assumed that the HL Tau disk is in a steady state (i.e., $\dot{M}=$ const.).
While this assumption may be reasonable in the current calculations,
it is intriguing to examine how the above results would be altered if this assumption is relaxed.
This is motivated by the observational fact that some YSOs undergo the so-called episodic accretion \citep[see below,][]{hhc16}.
To proceed, we re-compute the radius ($a$) of dust particles and the corresponding Stokes number ($\mbox{St}$) as done in Section \ref{sec:res}.
We here reduce the value of $\dot{M}$ from $10^{-7}$ M$_{\odot}$ yr$^{-1}$ to $5 \times 10^{-8}$ M$_{\odot}$ yr$^{-1}$.
Note that when $\dot{M}$ is increased by a factor of 2, even disk wind models cannot satisfy all the requirements simultaneously.
Figures \ref{fig8} and \ref{fig9} show the results: the former is for $a$ and the latter is for $\mbox{St}$.
On the left panel, the results are obtained, taking into account only turbulence
while turbulence and disk winds are both included on the right panel.
We find that if the accretion rate at the outer disk is lower than that at the inner disk by a factor of at least 2,
then both the disk models with and without the effect of disk winds can satisfy all the three requirements 
($H_{\rm d} \la 1$ au and $Q \ga 2$ at $r=100$, and $\mbox{St} \simeq 0.1$).
Our results also show that the reduction of $\dot{M}$ expands the parameter space considerably
in which disk wind models can reproduce the configuration of the HL Tau disk (see the right panel).
Based on this result, one may wonder whether disk winds may not be a necessary ingredient 
if the HL Tau disk is currently in non-steady accretion.
As discussed below, however, this is unlikely because the HL Tau configuration does not fit the basic properties of episodic accretion.
Thus, whereas the assumption of steady state accretion may not be fully valid,
our finding that disk winds are important can still hold.

In summary, it is obviously important to perform more detailed, numerical simulations to examine the validity of our results.
Nonetheless, it can be concluded that our main results may be maintained even if some assumptions are relaxed.

\subsection{Other mechanisms of disk accretion}

We finally discuss a possibility of how other mechanisms can serve as a plausible process to account for the key features of the HL Tau disk.
We here examine episodic accretion and GI.

We first explore the possibility of episodic accretion.
It is well known that some (perhaps most) YSOs experience accretion outbursts during the class I and II stages \citep[e.g.,][]{hhc16}.
There are two famous types \citep[e.g.,][]{aad14}: the one is the so-called FUor that is named from FU Ori \citep{h66,h77}, 
and the other is the EXor coined from EX Lupi \citep{h89}.
Both objects exhibit accretion outbursts, but the magnitude and the duration of the outbursts for FUors are much larger than those of EXors.
It is still inconclusive whether these two objects can be regarded as two distinct types and/or what would be the ultimate mechanism to trigger such outburst events.
It is, however, clear that the disk mass should be piled up during the quiescent phase,
and such masses are accreted onto the central star when an outburst occurs.
Since FUors generally have accretion rates higher than $10^{-5}$ M$_{\odot}$ yr$^{-1}$ during outbursts,
HL Tau may be better classified as the EXor if HL Tau is in the outburst phase.
Note that outbursts can increase the accretion rate up to $10^{-8}-10^{-6}$ M$_{\odot}$ yr$^{-1}$ for EXors.
Also, a quiescent phase is unlikely to be applicable for the current HL Tau system.
This is because, if this were the case, $\dot{M}$ in the outer disk should be larger than that in the inner disk ($\approx 10^{-7}$ M$_{\odot}$ yr$^{-1}$),
so that a pile up of masses are established in the inner disk.
We find that when $\dot{M} > 10^{-7}$ M$_{\odot}$ yr$^{-1}$ at the outer disk, GI would be the main driver of disk accretion there (see Section \ref{disc_limit}).
As discussed below, however, GI may be unlikely to reproduce the observed features of the HL Tau disk.
 
We now consider how the current configuration of the HL Tau system can be (in)consistent with the basic properties of Exors.
We begin with pointing out that EXor-like outbursts tend to occur on stars in the later stages of forming,
when the circumstellar envelope is already dispersed \citep{saa09,aad14}.
The disappearance of envelopes is obviously inconsistent with the HL Tau case.
Then we compute how much the disk mass can be piled up during the quiescent phase,
and how long an outburst event can last.
This calculation is motivated by the above results that if the accretion rate is $\la 5 \times 10^{-8}$ M$_{\odot}$ yr$^{-1}$ at the outer disk,
the MHD turbulence model can also reproduce the key features of the HL Tau disk (see the left panel of Figure \ref{fig9}).
Utilizing this value under the assumption that the accretion rate at the outer disk is comparable between the outburst and the quiescent phases,
the mass  ($\Delta m$) that can be accumulated during a quiescent phase ($\Delta t$) is written as
$\Delta m = \dot{M}_{\rm out} \Delta t$.
The observational results suggest that the duration of outbursts and the interval between them are about 1 yr and 10 yrs for EXors, respectively \citep{hk16,aad14}.
As a result, $\Delta m \la  5 \times 10^{-7}$ M$_{\odot}$($=5 \times 10^{-8}$ M$_{\odot}$ yr$^{-1} \times 10$ yr).
Furthermore, given that the current accretion rate is about $10^{-7}$ M$_{\odot}$ yr$^{-1}$,
we can estimate that an outburst should continue for $ \la 5$ yr, if HL Tau is currently in an outburst phase.
It is interesting that these numbers seem roughly consistent with the general properties of EXors.
This means that the turbulence model may be able to account for the observed properties of the HL Tau disk,
if HL Tau would be an outbursting EXor.
Finally, if HL Tau were an EXor, its outflows would show knots with spacing corresponding to the interval between outbursts.
However, no such features are reported for the molecular outflow \citep{kmmj16}.  
Thus, it seems unlikely that HL Tau is an outbursting system.

Next, we discuss the possibility of GI.
As already shown in Figure \ref{fig7},  a high value of the gas-to-dust ratio is needed for GI to operate in the HL Tau disk 
and to reproduce the observed high disk accretion rate.
In addition, it is difficult to achieve a thin dusty layer in self-gravitating disks.
This is because the root mean squared vertical turbulent speed is comparable to the sound speed in such disks \citep{sc14}.
Then the corresponding value of $\alpha_D$ becomes an order of unity (see equation (\ref{eq:alpha_D})), 
which suggests that $H_{\rm d} \simeq H_{\rm g}$ \citep[also see][]{bc16}.
Based on the above consideration, GI is unlikely to currently be the main driver of disk accretion for the HL Tau disk.

\section{Conclusions and summary} \label{sec:conc}

We have investigated a possibility of how magnetically driven disk accretion can account for a number of key features of the HL Tau disk.
Especially, we have focused on the recent ALMA observations which suggest a high degree of dust settling at the outer part of the disk.
This observational finding is interesting, because under the framework of the standard 1D viscous disk model, 
such a configuration is not necessarily compatible with other observational results 
which suggest that the accretion rate onto the central star is high.

We have developed a simplified, but physically motivated disk model in which the recent non-ideal MHD simulations are utilized.
Our model thereby can examine how the disk angular momentum is transported both radially via MHD turbulence and vertically via magnetically induced disk winds
for given values of the initial plasma beta ($\beta_0$) and of the sound speed ($c_{\rm s}$).
We find that these two features (a high disk accretion rate and a high degree of dust settling) can be reproduced well
when magnetically induced disk winds are properly taken into account (see Figure \ref{fig3}).
This becomes possible because disk winds can transport a significant amount of the disk angular momentum,
so that a high disk accretion rate is achieved without a higher level of turbulence (see Figure \ref{fig5}).
This naturally leads to efficient dust settling.
We have also discussed that the framework of magnetically driven disk accretion would be useful to investigate other features of the HL Tau disk
such as the global magnetic field configuration (see Figure \ref{fig6}) and the gas-to-dust ratio (see Figure \ref{fig7}).

Our results however show that the optimal configuration of the HL Tau disk is realized 
only in a very narrow parameter space under the assumption of steady accretion ($\beta_0 \simeq 2 \times 10^4$, see Figure \ref{fig4}).
In addition, our model heavily relies on the results of numerical simulations in which both the accretion and the wind stresses are vertically integrated (see Figure \ref{fig1}).
As other limitations of our model, we have discussed neglect of mass loss due to disk winds and the assumption of steady state accretion.
Higher spatial resolution observations for the disk gas and more self-consistent modeling would be required to fully identify the main driver of disk accretion for the HL Tau disk.
For instance, a coupling of dust growth and non-ideal MHD effects would be important 
to further explore the origin of the observed gaps in both the gas and dust distributions for the disk.
It would also be worth noticing that the recent non-ideal MHD simulations can generate gap structures in gas disks \citep[e.g.,][]{mo12,frd15,blf16,blf17,slk17}.
While it is out of the scope of this paper to identify the origin of dust gaps in the HL Tau disk,
these results are promising for magnetically driven disk accretion models.

Thus, understanding magnetically driven disk accretion would be the fundamental step to draw a better picture of disk evolution and the subsequent planet formation there,
and the HL Tau disk is one of the paramount examples to perform such investigations.

%% The \notetoeditor{TEXT} command allows the author to communicate
%% information to the copy editor.  This information will appear as a
%% footnote on the printed copy for the manuscript style file.  Nothing will
%% appear on the printed copy if the preprint or
%% preprint2 style files are used.

\acknowledgments

The authors thank Andrea Isella for stimulating discussions,
and an anonymous referee for useful comments on our manuscript.
This research was carried out at Jet Propulsion Laboratory, California Institute of Technology under a contract with NASA.
Work on this project by Y.H. and N.J.T. was supported by JPL/Caltech.
S.~O. is supported by Grants-in-Aid for Scientific Research
(\#15H02065, 16K17661, 16H04081) from MEXT of Japan.
Work by M.F. was funded by NASA Exoplanet Research Program grant 14-XRP14\_2-0153.




%\clearpage

%\appendix

\bibliographystyle{apj}          

\bibliography{apj-jour,adsbibliography}    %% includes the journal abbrevs

\end{document}